\documentclass[aps, prb, reprint, superscriptaddress, amsmath, amssymb, showkeys]{revtex4-2}

\usepackage{graphicx}
\usepackage{dcolumn}
\usepackage{bm}
\usepackage{soul}
\usepackage{color}
\usepackage{amsmath}
\usepackage{mathtools}
\usepackage[colorlinks=true, allcolors=blue]{hyperref}

\DeclareUnicodeCharacter{2061}{}

\begin{document}

\title{Ferrimagnetic hexagonal Mn$_2$CuGe Heusler alloy\\with a low-temperature spin-glass state}

\author{Abhinav Kumar Khorwal}
\affiliation{Department of Physics, Central University of Rajasthan, Ajmer - 305817, Rajasthan, India}

\author{Sonu Vishvakarma}
\affiliation{Department of Physics, Pranveer Singh Institute of Technology, Kanpur - 209305, Uttar Pradesh, India}
\affiliation{Department of Physics, Indian Institute of Technology Madras, Chennai - 600036, Tamil Nadu, India}

\author{Sujoy Saha}
\affiliation{Department of Physics, Central University of Rajasthan, Ajmer - 305817, Rajasthan, India}

\author{Debashish Patra}
\affiliation{Department of Physics, Indian Institute of Technology Madras, Chennai - 600036, Tamil Nadu, India}

\author{Akriti Singh}
\affiliation{Department of Physics, Indian Institute of Science Education and Research Bhopal, Bhopal – 462066, Madhya Pradesh, India}

\author{Surajit Saha}
\affiliation{Department of Physics, Indian Institute of Science Education and Research Bhopal, Bhopal – 462066, Madhya Pradesh, India}

\author{V. Srinivas}
\affiliation{Department of Physics, Indian Institute of Technology Madras, Chennai - 600036, Tamil Nadu, India}

\author{Ajit K. Patra}
\email[Corresponding author: ]{a.patra@curaj.ac.in}
\affiliation{Department of Physics, Central University of Rajasthan, Ajmer - 305817, Rajasthan, India}

\begin{abstract}
An extensive experimental investigation on the structural, static magnetic, and non-equilibrium dynamical properties of polycrystalline Mn$_2$CuGe Heusler alloy using powder X-ray diffraction, DC magnetization, magnetic relaxation, magnetic memory effect, and specific heat measurements is presented. Structural studies reveal that the alloy crystallizes in a mixed hexagonal crystal structure (space groups P3c1 (no. 158) and P6$_3$/mmc (no. 194)) with lattice parameters a = b = 7.18(4) $\AA$ and c = 13.12(4) $\AA$ for the majority phase. The DC magnetization analysis reveals a paramagnetic to ferrimagnetic phase transition around T$_C$ $\approx$ 682 K with a compensation of magnetization at $\approx$ 250 K, and a spin-glass transition around T$_P$ $\approx$ 25.6 K. The N\'eel theory of ferrimagnets supports the ferrimagnetic nature of the studied alloy and the estimated T$_C$ ($\approx$ 687 K) from this theory is consistent with that obtained from the DC magnetization data. A detailed study of non-equilibrium spin dynamics via magnetic relaxation and memory effect experiments shows the evolution of the system through a number of intermediate states and striking magnetic memory effect. Furthermore, heat capacity measurements suggest a large electronic contribution to the specific heat capacity suggesting strong spin fluctuations, due to competing magnetic interactions. All the observations render a spin-glass behavior in Mn$_2$CuGe, attributed to the magnetic frustration possibly arising out of the competing ferromagnetic and antiferromagnetic interactions.
\end{abstract}


\maketitle

\section{\label{sec:Introduction}Introduction}

The field of spintronics is a rapidly developing extension of conventional electronics, that incorporates the additional spin degree of freedom along with the charge of electrons, which enables in designing novel devices with high throughput, enhanced power-efficiency, increased storage density, etc. \cite{Ref1}. These spintronic devices mostly utilize ferromagnetic (FM) materials as an active component. However recently, ferrimagnetic (FiM) and antiferromagnetic (AFM) spintronics have gained tremendous attention for realizing efficient memory and data storage applications. The nearly/fully compensated magnetic moments in FiM and AFM materials help in minimizing stray magnetic fields and ensure robust device performance under external fields, thus reducing energy losses. Therefore, from a technological point of view, these materials are advantageous over conventionally used ferromagnets \cite{Ref2, Ref2a, Ref2b, Ref2c}. FiM materials are superior to AFM materials for spintronics because they provide combined benefits of both ferromagnets and antiferromagnets such as faster spin dynamics than ferromagnets, and prove to be promising for high-density spintronic devices \cite{Ref2}. Moreover, their asymmetric sub-lattices with distinct material properties lead to interesting phenomenon such as perfectly spin-polarized currents, an essential requirement in spintronics \cite{Ref2, Ref3}. Hence, developing new FiM materials with high T$_C$, for practical applications, is of utmost significance in current materials research.

Heusler alloys (HAs), a fascinating class of materials, provide an exceptional playground for designing novel materials with diverse and tunable physical properties such as ferromagnetism, antiferromagnetism, ferrimagnetism, half-metallicity (HM), shape memory effect, topological insulating behavior, magnetoresistance, magnetocaloric effect, etc. \cite{Ref4, Ref5, Ref5a}.

Since the last decade, Mn-based HAs have received much attention for their potential applications as HM FiM materials. HM materials, due to their 100\% spin polarization and enhanced magnetoresistance, offer tremendous advantages in advanced spintronic applications. Several Mn-based Heusler alloys show a HM band structure with high Curie temperatures and hence, are promising candidates for advanced applications \cite{Ref6, Ref7, Ref8, Ref9, Ref10}. HM is predicted using band structure calculations in Mn$_2$VZ (Z = Al, Si) \cite{Ref6, Ref7, Ref11}, Mn$_2$FeZ (Z = Al, Sb) \cite{Ref9, Ref11}, Mn$_2$CoZ (Z = Al, Si, Ge, Sn, Sb, Ga) \cite{Ref12, Ref10, Ref11, Ref13}, Mn$_2$YAl (Y = Cr, Mn) \cite{Ref11, Ref8}, Mn$_2$CrSb \cite{Ref8} and Mn$_2$RhSi \cite{Ref14}. However, only Mn$_2$CoSb \cite{Ref12} is experimentally reported suggesting its HM behavior.

Moreover, many Mn-based Heusler alloys are also found to possess FiM behavior, in which the magnetic moments of the constituent sub-lattices align in opposite directions. Density functional theory (DFT) predicts the FiM ground state for Mn$_2$VZ (Z = Al, Si, In) \cite{Ref6, Ref7, Ref15}, Mn$_2$RhSi \cite{Ref14} and Mn$_2$PtGa \cite{Ref16}. Further, Mn$_2$FeGa \cite{Ref17}, Mn$_2$PtZ (Z = Ga, In) \cite{Ref16, Ref18} are experimentally found to be FiM in nature, while Mn$_2$PtAl shows a nearly compensated FiM behavior \cite{Ref19}. In order to achieve the fully compensated FiM state, elemental substitution as well as composition variation in Mn$_2$Co$_{0.5}$V$_{0.5}$Z (Z = Al, Ga) \cite{Ref20, Ref21} and Mn$_{1.5}$FeV$_{0.5}$Al \cite{Ref22} are also explored experimentally.

Unlike other 3d elements such as Fe and Co, the Cu atom has the additional 4s sub-shell. Since electronic and magnetic properties are heavily influenced by the electronic configuration, interesting electronic and magnetic properties are expected in Mn$_2$CuZ HAs due to the presence of the additional 4s sub-shell of Cu, as compared to other Mn-based alloys \cite{Ref23}.

The electronic structure calculations reveal that Mn$_2$CuAl possesses a FiM ground state and with a small contraction of the equilibrium lattice, it becomes a HM antiferromagnet \cite{Ref24}. Experimentally prepared Mn$_2$CuAl ribbons show a FiM transition at 690 K with a compensation point at 630 K \cite{Ref25}. Several HM FiM HAs in Mn$_2$CuZ (Z = Mg, Sb, Ge) systems \cite{Ref26, Ref27, Ref28, Ref23} are recently predicted using DFT. A FiM ground state for Mn$_2$CuGe with CuHg$_2$Ti-type (inverse Heusler) crystal structure is predicted. Additionally, a HM behavior with 100\% spin polarization at the Fermi level is also anticipated for this system \cite{Ref28}. It is further explored using first-principle calculations that the HM FiM behavior of Mn$_2$CuGe is insensitive to the expansion of lattice parameters (by $\sim$ 2.5 \%) indicating its robust magnetic ground state \cite{Ref23}. In CuHg$_2$Ti-type crystal structure, the Mn(A) and Mn(B) sub-lattices are antiparallelly aligned to each other with magnetic moments of -1.44 $\mu_B$ and 2.44 $\mu_B$, respectively. Cu contributes a small magnetic moment (-0.06 $\mu_B$), leading to a low total magnetic moment of 0.97 $\mu_B/f.u.$, that is in good agreement with Slater-Pauling rule \cite{Ref23}. Although there are a few theoretical reports predicting several functional properties of Mn$_2$CuGe HA, there is no report available on any experimental investigation on this material. Considering the importance of FiM materials on the development of advanced spintronics-based technology and the promising properties predicted for Mn$_2$CuGe, a thorough experimental investigation of this alloy is highly desirable.

In the present study, a detailed investigation of the structural and magnetic properties of polycrystalline Mn$_2$CuGe HA is presented. At room temperature, the arc-melted  Mn$_2$CuGe HA crystallizes in a hexagonal structure with a minor impurity phase. The magnetization study reveals a PM - FiM transtion at $\sim$ 682.8 K with a compensation point at $\sim$ 250 K. The low temperature magnetic features such as bifurcation in ZFC and FC magnetization with a peak temperature T$_P$, unidirectional exchange-bias, magnetic memory and relaxation below T$_P$, along with heat capacity analysis indicate the presence of low temperature spin-glass (SG) phenomenon in Mn$_2$CuGe which may be attributed to the presence of competing magnetic interactions.

\section{\label{sec:Experimental details}Experimental details}

Polycrystalline Mn$_2$CuGe samples are synthesized by conventional arc-melting technique using high-purity elemental constituents in stoichiometric amounts under an argon atmosphere. The elements (Mn, Cu and Ge) are of high purity (99.99 \%) and obtained from Alfa Aesar. The samples are remelted several times to promote homogeneity and the ingots are annealed at 1000 $^\circ$C for 72 hours followed by water quenching. The structural and elemental composition of the samples are carefully examined using a Rigaku X-ray diffractometer with Cu-K$_{\alpha}$ radiation ($\lambda$ = 1.5406 $\AA$) and a field emission scanning electron microscope (Fe-SEM) (Inspect F) equipped with energy-dispersive X-ray (EDX) spectroscopy, respectively. The chemical compositions are measured at multiple spots on the sample surface to confirm their chemical homogeneity. The DC magnetization measurements are carried out using vibrating sample magnetometer (VSM) attachments to a Quantum Design physical property measurement system (PPMS, DynaCool) and to a superconducting quantum interference device (SQUID) in Quantum Design magnetic property measurement system (MPMS 3). The specific heat (C$_p$) is measured using a Quantum Design physical property measurement system (PPMS) in the temperature range of 1.8 K to 200 K adopting the relaxation technique in a zero magnetic field.

\section{\label{sec:Results and discussion}Results and discussion}

\subsection{\label{sec:Structure and composition}Structure and composition}

\begin{figure}
    \includegraphics[width=0.45\textwidth]{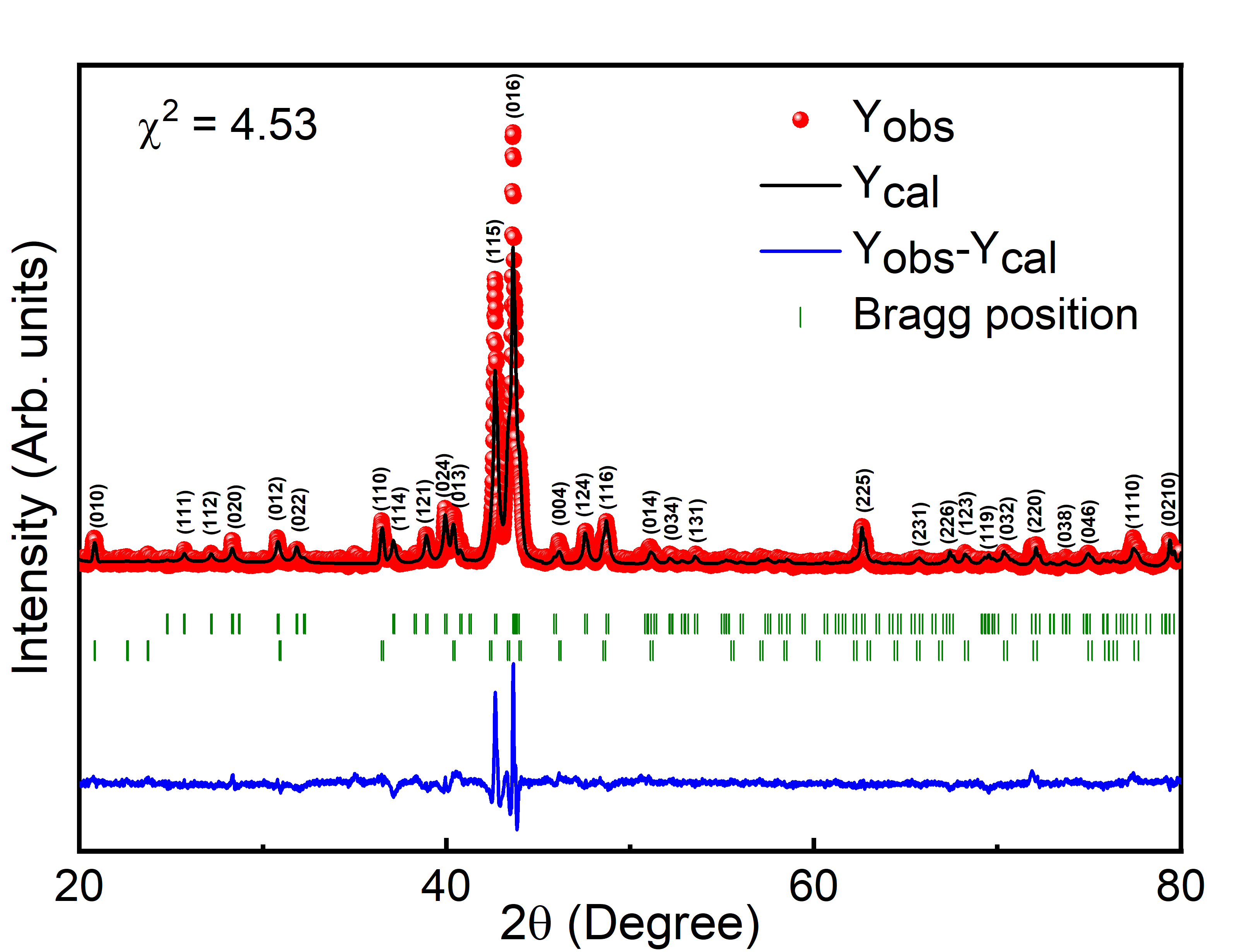}
    \caption{\label{fig:XRD}Rietveld refinement of the X-ray diffraction pattern of Mn$_2$CuGe at room temperature.  The solid red spheres and solid black line are the observed and calculated patterns, respectively. The Bragg positions are indicated by green ticks. The solid blue line at the bottom represents the difference between the observed and calculated intensities. $\chi^2$ = 4.53 represents the goodness of fit.}
\end{figure}

In order to examine the phase purity and crystal structure of the prepared alloy, powder X-ray diffraction (XRD) is performed at room temperature. Fig. \ref{fig:XRD} displays the room temperature powder XRD data of Mn$_2$CuGe HA from 20$^\circ$ to 80$^\circ$ along with Rietveld refinement using the FullProf software. Rietveld refinement shows that the sample crystallizes in a mixture of two hexagonal phases: Mn$_5$Ge$_2$-type crystal structure (space group P3c1 (no. 158)) with lattice parameters a = b = 7.18(4) $\AA$, c=13.12(4) $\AA$, $\alpha = \beta = 90^\circ$, $\gamma=120^\circ$, and Cu$_{1.5}$Ge$_{0.5}$Mn-type crystal structure (space group P6$_3$/mmc (no. 194)) with lattice parameters a = b = 4.92(5) $\AA$, c = 7.86(6) $\AA$, $\alpha = \beta = 90^\circ$, $\gamma=120^\circ$. The solid red spheres and solid black lines are the observed and the calculated patterns, respectively. The Bragg positions are indicated by green ticks. The solid blue line at the bottom represents the difference between the observed and the calculated intensities (Fig. \ref{fig:XRD}). The obtained goodness of fit is $\chi^2$ = 4.53.  The fraction of the impurity phase (P6$_3$/mmc) is found to be 7.95 wt\% using the dual-phase Rietveld refinement process. The phase details, lattice parameters, unit-cell volume, and other fitting parameters derived from Rietveld refinement are listed in Table \ref{tab:XRD}.

\begin{figure}
    \includegraphics[width=0.45\textwidth]{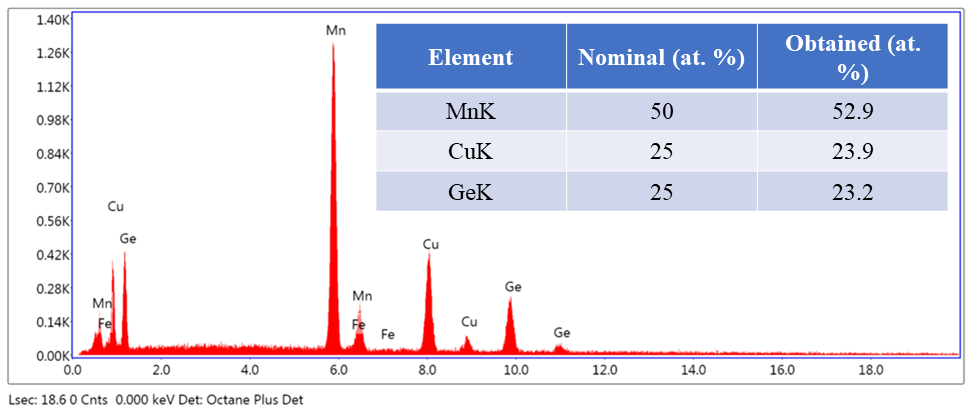}
    \caption{\label{fig:EDX}EDX spectra of Mn$_2$CuGe compound Inset: Nominal and average elemental composition}
\end{figure}

The surface micro-structure is characterized by using FE-SEM on a rectangular piece of polished alloy ingot. The elemental analysis is carried out using EDX spectroscopy attached to a FE-SEM at different regions of the sample in order to confirm its chemical homogeneity. The EDX spectroscopy analysis reveals that the alloy compositions are in good agreement with its nominal composition (Fig. \ref{fig:EDX}).

\subsection{\label{sec:Magnetic properties}Magnetic properties}

\begin{figure}
    \includegraphics[width=0.45\textwidth]{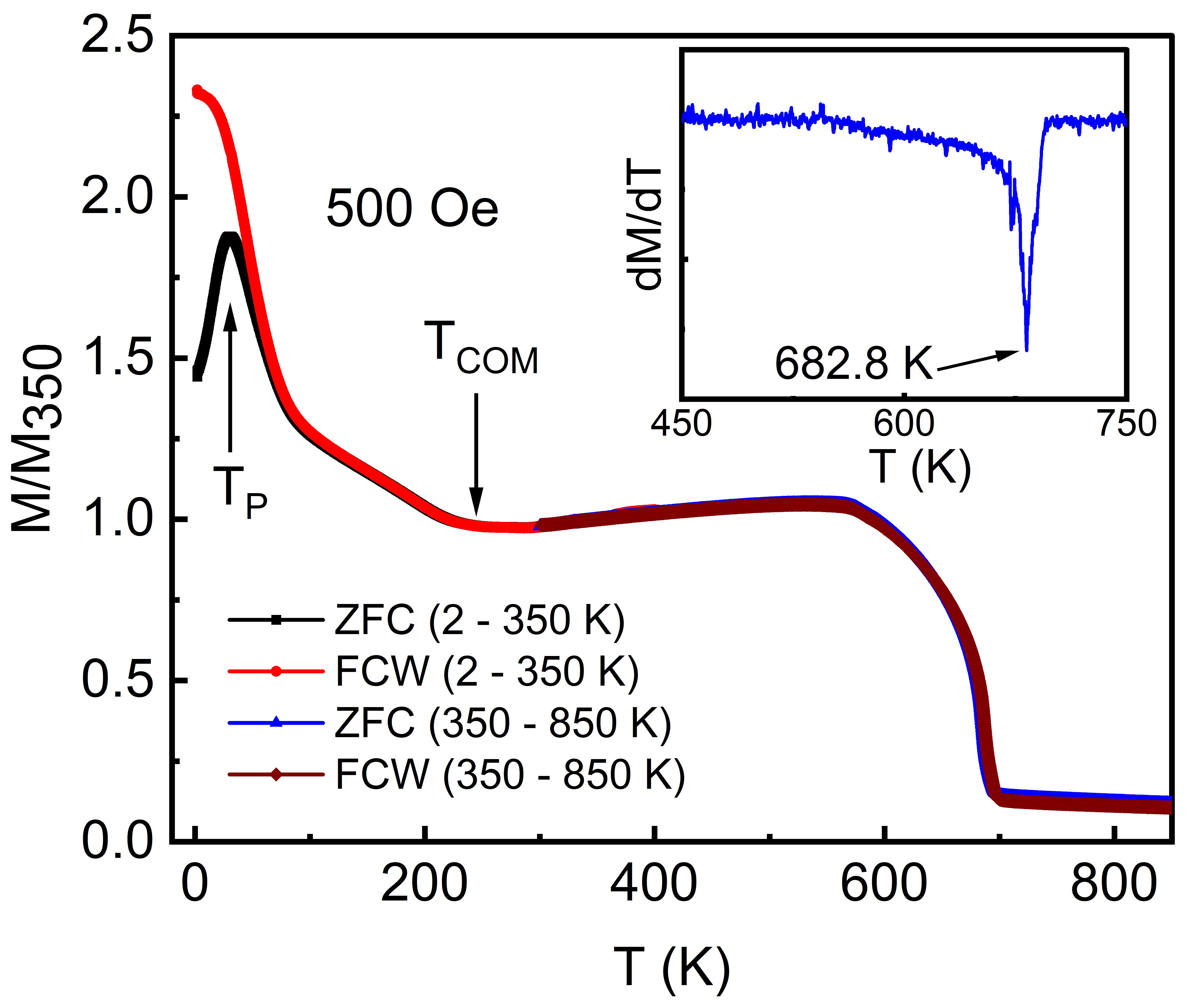}
    \caption{\label{fig:MT}Scaled magnetization as a function of temperature measured at 500 Oe in the temperature range of 2 K - 850 K for Mn$_2$CuGe alloy. Inset shows the temperature dependent first derivative of FCW magnetization with respect to the temperature (dM$_{FCW}$/dT).}
\end{figure}

Fig. \ref{fig:MT} presents the scaled magnetization data in  zero-field-cooled (ZFC) and field-cooled warming (FCW) modes as a function of temperature measured at an applied magnetic field of 500 Oe in the temperature range of 2 K - 850 K. In the ZFC curve, with decreasing temperature the magnetization shows an abrupt jump at T = 682.8 K (obtained from the minimum of first derivative of FCW magnetization as a function of temperature; inset of Fig. \ref{fig:MT}) which can be ascribed to the PM - FiM transition temperature (T$_C$). On further decreasing the temperature, magnetization shows a convex behavior with a compensation of magnetization at T$_{COM}$ $\approx$ 250 K and keeps on increasing to form a peak around $\sim$ 25.6 K (denoted as T$_P$). Similar compensation of magnetization is also observed for other compensated FiM HAs \cite{Ref20, Ref21, Ref22}. The FC magnetization curve traces the ZFC curve up to a temperature of $\sim$ 200 K, below which it bifurcates and keeps on increasing up to the lowest temperature. The observed bifurcation of ZFC and FCW curves, and a peak in the ZFC curve at low temperature are characteristic signatures of SGs. The bifurcation of the ZFC and FCW curves originates from the spontaneous uncompensated spins which get locked at low temperatures \cite{Ref7, Ref29, Ref30}.
\begin{table*}
    \caption{\label{tab:XRD}Refined parameters for both the phases (lattice parameters, volume, RF-factor, Bragg-factor and fraction of phase).}
    \begin{ruledtabular}
        \begin{tabular}{cccccc}
            Phase (Space group, Number) & Lattice parameters (\AA) & Volume (\AA)$^3$ & RF-factor & Bragg factor & Fraction of phase\\
            \hline
            Hexagonal (P3c1, 158) & a = b = 7.18(4), c = 13.12(4) & 586.65 & 17.5 & 16.6 & 92.05\\
            Hexagonal (P6$_3$/mmc, 194) & a = b = 4.92(5), c = 7.86(6) & 165.27 & 12.2 & 17.3 & 7.95\\
        \end{tabular}
    \end{ruledtabular}
\end{table*}

\begin{figure}
    \includegraphics[width=0.45\textwidth]{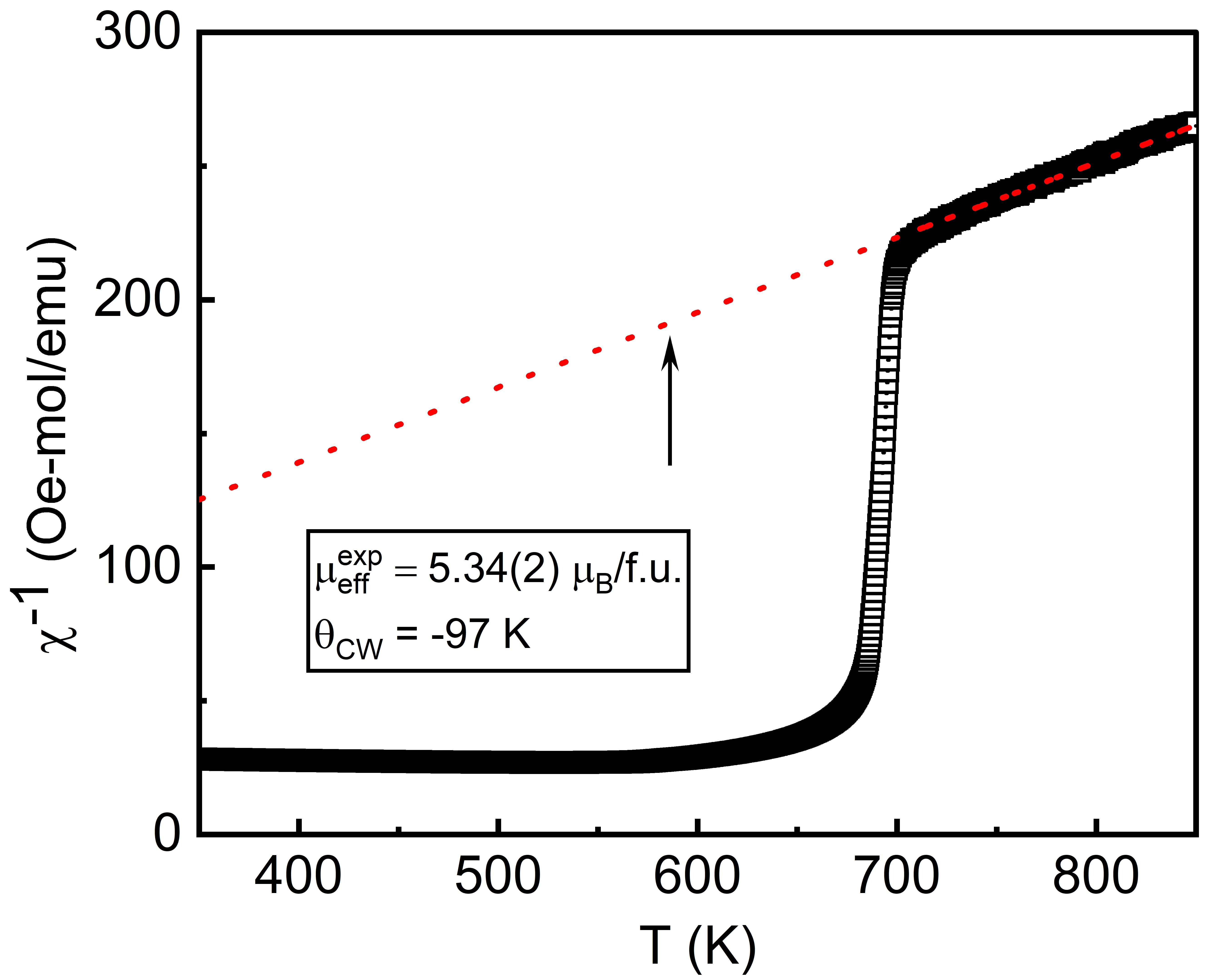}
    \caption{\label{fig:CW}Inverse susceptibility versus temperature along with Curie-Weiss fit (dotted red line) in the temperature range of 710 K - 850 K at 500 Oe applied magnetic field.}
\end{figure}

The temperature variation of inverse susceptibility ($\chi^{-1}$ = H/M) at 500 Oe is presented in Fig. \ref{fig:CW}. The plot of $\chi^{-1}$ shows a step like behavior. In the paramagnetic region (T $>$ 710 K), $\chi^{-1}$ follows a linear  temperature dependency and it is analyzed using the Curie-Weiss law \cite{Ref29} given as,
\begin{equation}
    \chi(T) = \frac{C}{T-\theta_{CW}},
    \label{eqn:1}
\end{equation}
where C is the Curie constant and $\theta_{CW}$ is the Curie-Weiss temperature. The least square fit to equation (\ref{eqn:1}) in the temperature range of 710 K - 850 K gives the value of C = 3.57(4) emu K mol$^{-1}$ Oe$^{-1}$ with a negative value of $\theta_{CW}$ (- 97 K). The temperature variation of magnetization shows a FiM nature of the prepared alloy whereas a negative $\theta_{CW}$ indicates strong antiferromagnetic interactions between the sub-lattices.

\begin{figure}
    \includegraphics[width=0.45\textwidth]{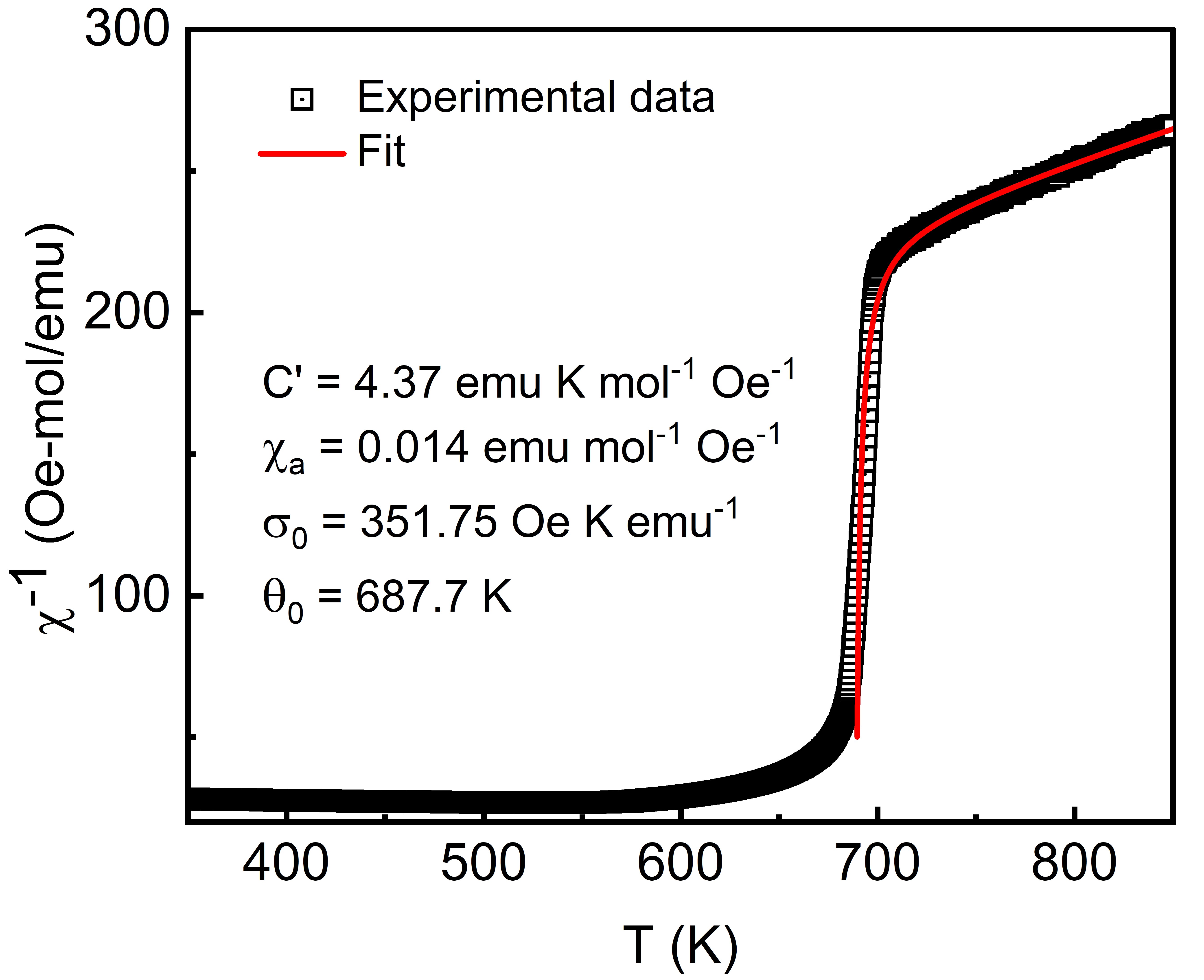}
    \caption{\label{fig:Neel}Temperature-dependent inverse susceptibility at 500 Oe magnetic field. The solid line indicates the least square fit to N\'{e}el expression of ferrimagnets (equation \ref{eqn:2}) at T $>$ T$_C$.}
\end{figure}

Both theoretical studies \cite{Ref23, Ref28} and the thermo-magnetization measurement suggest the FiM ground state of Mn$_2$CuGe. Hence, to have a deeper understanding about the FiM nature of Mn$_2$CuGe, the deviation in $\chi^{-1}$ from a linear behavior versus temperature is analyzed at T $>$ T$_C$ for 500 Oe applied magnetic field using the N\'{e}el expression for FiM \cite{Ref31, Ref32}, which is described by following equation:
\begin{equation}
    \frac{1}{\chi} = \frac{T}{C'} + \frac{1}{\chi_a} + \frac{\sigma_0}{(T-\theta_0)},
    \label{eqn:2}
\end{equation}
where C$'$ denotes the Curie constant in the absence of any interactions and $\chi_a$, $\theta_0$ and $\sigma_0$ have functional forms which relates to the molecular field constants (N$_{AA}$, N$_{BB}$ and N$_{AB}$) for two sub-lattice model of FiM with sub-lattice A and B \cite{Ref32}. The Curie constants related to the sub-lattices A and B are C$_A$ and C$_B$, respectively and can be calculated using C$'$ = C$_A$ + C$_B$. Fig. \ref{fig:Neel} displays the temperature-dependent inverse susceptibility along with the fitting (solid red line) to equation (\ref{eqn:2}). The obtained values of the parameters are C$'$ = 4.37 emu K mol$^{-1}$ Oe$^{-1}$, $\chi_a$ = 0.014 emu mol$^{-1}$ Oe$^{-1}$, $\sigma_0$ = 351.75 Oe K emu$^{-1}$ and $\theta_0$ = 687.7 K. The intra-sub-lattice and inter-sub-lattice molecular field constants of sub-lattices A and B namely N$_{AA}$, N$_{BB}$ and N$_{AB}$ can be calculated using the following equations \cite{Ref33}:
\begin{equation}
    \frac{1}{\chi_a} = \frac{1}{C'^2} (C_A^2 N_{AA} + C_B^2 N_{BB} + 2C_AC_BN_{AB}),
    \label{eqn:3}
\end{equation}
\begin{equation}
    \theta_0 = \frac{C_AC_B}{C'} (2N_{AB}-N_{AA}-N_{BB}),
    \label{eqn:4}
\end{equation}
\begin{align}
    \sigma_0 = \frac{C_AC_B}{C'^2} [C_A^2 (N_{AA}-N_{AB})^2 + C_B^2 (N_{BB}-N_{AB})^2 \nonumber\\ - 2C_AC_B\{N_{AB}^2 - (N_{AA} + N_{BB})N_{AB} + N_{AA}N_{BB}\}].
    \label{eqn:5}
\end{align}

Using the Curie constant C$'$ = C$_A$ + C$_B$ and C$_A$/C$_B$ = 1.38 \cite{Ref33}, the values of Curie constants C$_A$ and C$_B$ corresponding to the A and B sub-lattices, respectively, are calculated as C$_A$ = 2.53 emu K mol$^{-1}$ Oe$^{-1}$ and  C$_B$ = 1.84 emu K mol$^{-1}$ Oe$^{-1}$. These values are then used in equations (\ref{eqn:3}), (\ref{eqn:4}) and (\ref{eqn:5}) to obtain the molecular field constants N$_{AA}$, N$_{BB}$ and N$_{AB}$. Since equation (\ref{eqn:5}) is quadratic in nature, there are two sets of solutions obtained. Set I: N$_{AA}$ = - 48.75, N$_{BB}$ = - 137.15 and N$_{AB}$ = 229.97, and Set II: N$_{AA}$ = - 37.48, N$_{BB}$ = - 152.64 and N$_{AB}$ = 227.86. According to Ehrenberg et al. \cite{Ref34}, the occurrence of T$_{COM}$ requires that the intra-sub-lattice exchange constants N$_{AA}$ and N$_{BB}$ be sufficiently different. This is true for both the obtained sets of solutions. Hence, both sets are acceptable solutions. From the obtained molecular field constants N$_{AB}$ = n, N$_{AA}$ = n$\eta$, N$_{BB}$ = n$\xi$, where $\eta$ and $\xi$ are the ratio of A-A and B-B interactions to A-B interaction, respectively. Using Set I, the values of $\eta$ = - 0.21, $\xi$ = - 0.60 and n = 229.97, while Set II gives the values $\eta$ = -0 .16, $\xi$ = -0 .67 and n = 227.86. The negative values of $\eta$ and $\xi$ are due to the competing FM - AFM interactions, and similar values ($\eta$ = - 0.24, $\xi$ = - 0.20) are reported for Mn$_2$Sb \cite{Ref31}. Further, the obtained values of $\eta$, $\xi$, n, C$'$, C$_A$ and C$_B$ are used to estimate T$_C$ by the following equation \cite{Ref32, Ref35}:
\begin{equation}
    T_C = \frac{n}{2} \left [ -(\eta C_A + \xi C_B) + \{(\eta C_A - \xi C_B)^2 + 4C_AC_B\} \right].
    \label{eqn:6}
\end{equation}

The estimated value of T$_C$ $\sim$ 687 K is in good agreement with the T$_C$ obtained from the temperature-dependent magnetization (T$_C$ = 682.8 K). The analysis of molecular field constants also supports the FiM nature of Mn$_2$CuGe HA that is in excellent agreement with the theoretical predictions.

\begin{figure}
    \includegraphics[width=0.45\textwidth]{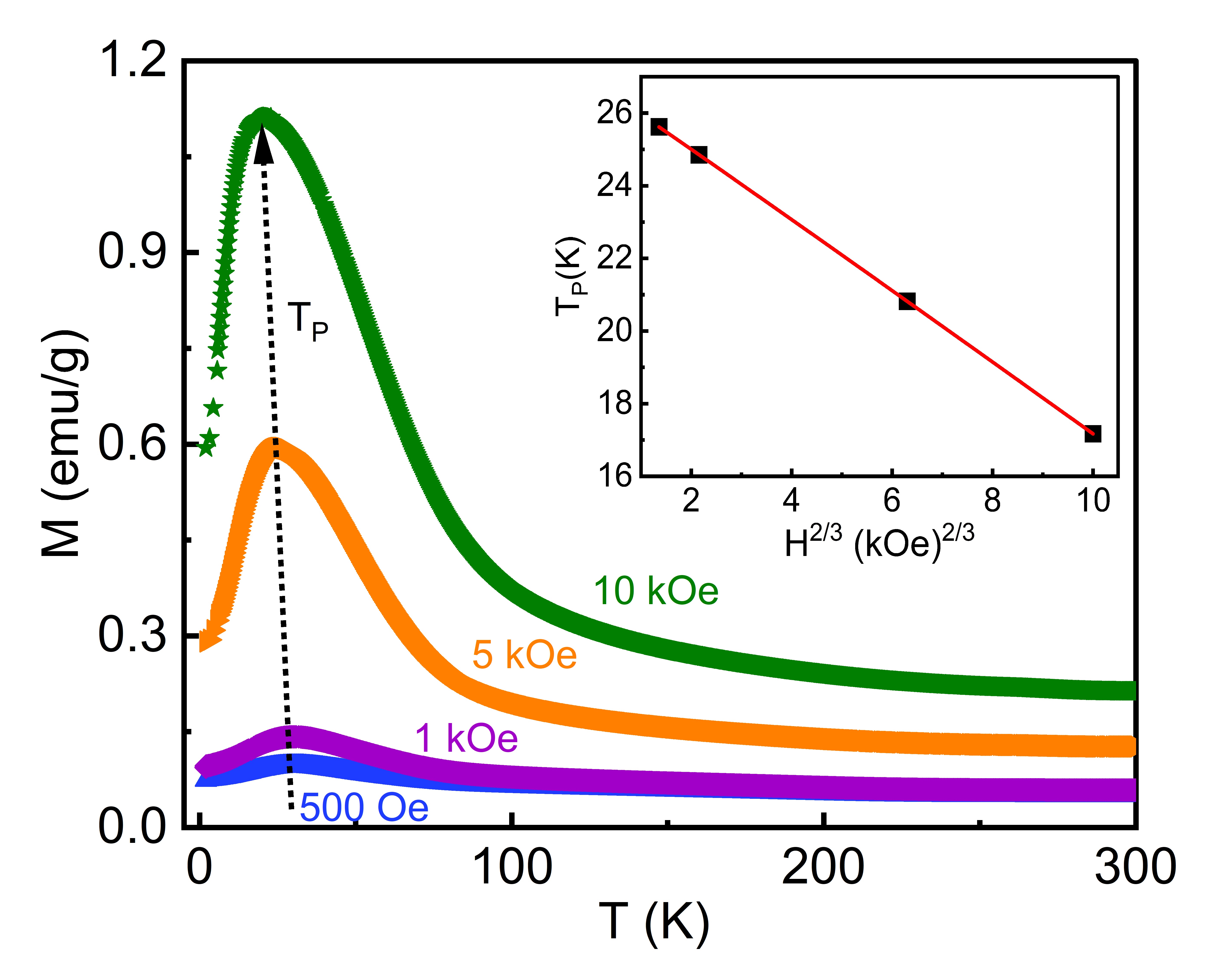}
    \caption{\label{fig:MT_combined}Temperature-dependent magnetization curves of Mn$_2$CuGe alloy in ZFC and FC modes measured at Various applied magnetic fields (0.5, 1, 5 and 10 kOe) in the temperature range of 2 K - 300 K. Inset represents the field dependence of peak temperatures (T$_P$).}
\end{figure}

In order to understand the low temperature magnetic anomaly, M(T) at various applied magnetic fields is measured (see Fig. \ref{fig:MT_combined}). The observed peak temperature T$_P$ $\approx$ 25.6 K in ZFC curve at 500 Oe shifts to lower temperature with increasing magnetic fields ($\sim$ 17.17 K at 10 kOe). T$_P$ is plotted as a function of H$^{2/3}$ (as shown in the inset of Fig. \ref{fig:MT_combined}) given by de Almeida-Thouless \cite{Ref30, Ref36} for low magnetic field range which is well known as the AT line. The T$_P$ at different fields perfectly obey the AT line, suggesting a strong irreversibility in magnetization. This type of T-H behavior are reported for other SG systems as well \cite{Ref37, Ref38}.
\begin{figure}
    \includegraphics[width=0.45\textwidth]{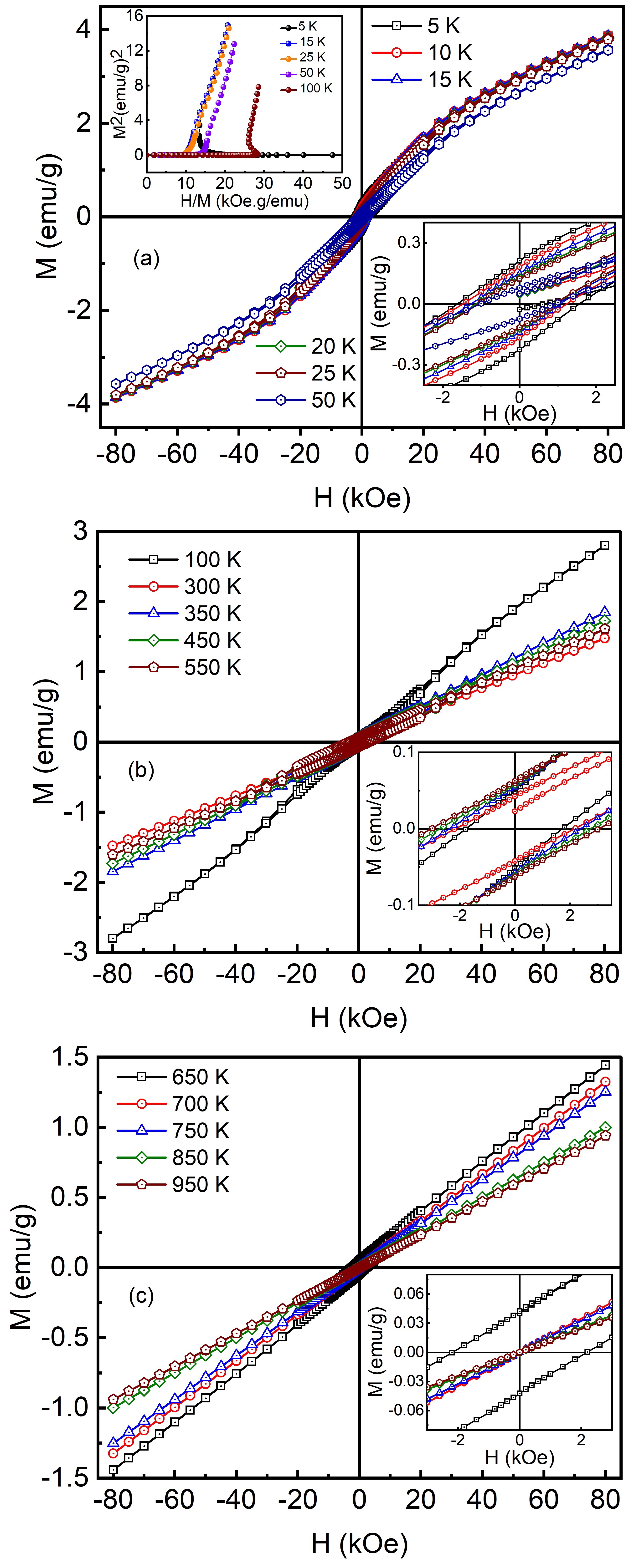}
    \caption{\label{fig:MH}(a): M-H hysteresis curves (5 K - 50 K). The curves exhibit a non-saturating small hysteresis with non-zero coercivity and remanence indicating competing AFM and FM interactions. Lower inset: Enlarged view near zero-field region. Coercivity (H$_C$) gradually decreases with increasing temperature. Upper inset: M$^2$ vs H/M curve (Arrott plot) suggesting the absence of spontaneous magnetization. (b): In the range 100 K - 550 K, H$_C$ increases with increasing temperature. (c): H$_C$ decreases and M-H curves become almost linear above 700 K.}
\end{figure}

To understand more about the low-temperature glassy phase, magnetization (M) as a function of applied magnetic field (H) in the zero-field-cooled mode are recorded at selected temperatures between 5 K and 950 K as shown in Fig. \ref{fig:MH}. The field dependent magnetization of Mn$_2$CuGe HA shows a non-saturating behavior even at a very high applied magnetic field of $\pm$ 80 kOe. However, another fact as observed from the M-H hysteresis loops is that at all the temperatures, the sample exhibits non-zero coercivity and remanence (shown in the lower insets of Fig. \ref{fig:MH}\hyperref[fig:MH]{(a-c)}). This observation may be attributed to the presence of a weak ferromagnetic component which is superimposed on the non-saturating linear AFM component. Such a feature corresponds to the presence of competing FM - AFM interactions, and is another important aspect in the creation of the SG state \cite{Ref39}. The observed highest coercive field (H$_C$ $\approx$ 3 kOe) is achieved at 550 K. Similar values of coercivity are also reported for other Mn-based Heusler alloys \cite{Ref16, Ref17, Ref19}. In the upper inset of Fig. \ref{fig:MH}\hyperref[fig:MH]{(a)} no positive y-intercept is observed in the Arrott plot (M$^2$ vs H/M). Therefore, the presence of any long-range ordering or spontaneous magnetization above T$_P$ is ruled out. The analysis of M(T) and M(H) data, thus suggests that Mn$_2$CuGe HA shows typical signatures of a compensated ferrimagnet, that is in excellent agreement with the theoretical predictions \cite{Ref23, Ref28}.

\subsection{\label{sec:Spin-glass behavior at low temperature}Spin-glass behavior at low temperature}

\subsubsection{\label{sec:Unidirectional exchange-bias effect}Unidirectional exchange-bias effect}

\begin{figure}
    \includegraphics[width=0.45\textwidth]{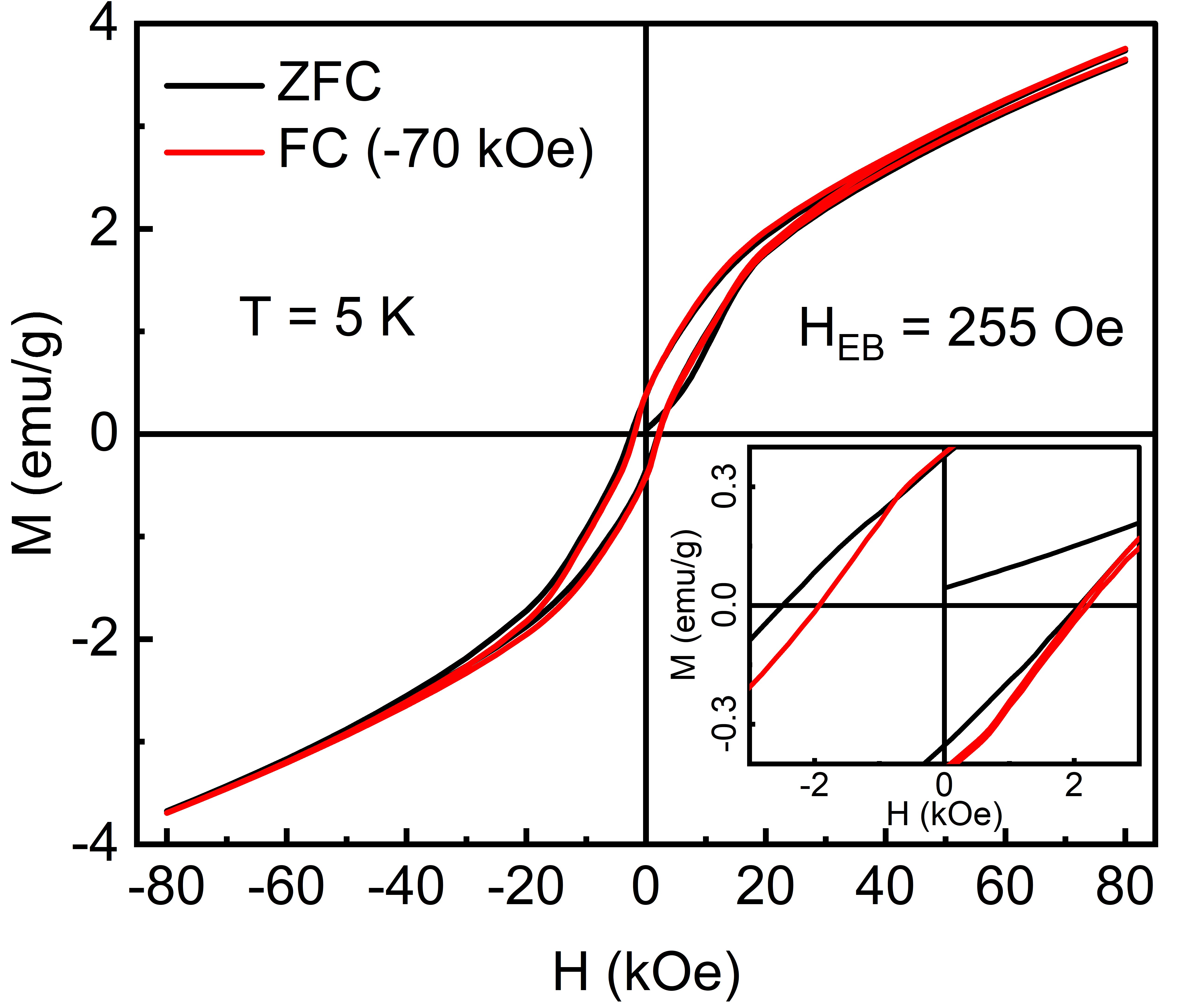}
    \caption{\label{fig:EB}M-H hystersis loop of Mn$_2$CuGe HA recorded at 5 K in ZFC and FC (- 70 kOe) protocols. Inset represents the enlarged image of M-H loop near the zero-field region to show exchange-bias effect.}
\end{figure}

There are several reports on the unidirectional exchange-bias effect in Mn-based HAs \cite{Ref16, Ref17, Ref19, Ref39}. The exchange-bias effect is mainly dictated by competing interactions at the interface of two opposite magnetically ordered lattices (FM and AFM sublattices) due to mixed magnetic phases and inhomogeneity in the material \cite{Ref19}. Furthermore, anisotropy plays a vital role for the exchange-bias phenomenon. The unidirectional exchange-bias is the shifting of FC hystersis curve in the opposite direction to that of the cooling field from the ZFC virgin curve. To check this effect, the prepared sample is cooled from 300 K to 5 K in ZFC and FC (- 70 kOe) and the magnetic field dependent isothermal magnetization is recorded (shown in Fig. \ref{fig:EB}). It can be seen from Fig. \ref{fig:EB}, that the FC hystersis loop is shifted to the positive field direction with an exchange-bias field of $\sim$ 255 Oe. Similar unidirectional exchange-bias effect is observed for other Mn-based FiM HAs such as, Mn$_2$PtGa (1.6 kOe) \cite{Ref16}, Mn$_2$FeGa (1.32 kOe) \cite{Ref17}, and Mn$_2$PtAl (2.73 kOe) \cite{Ref19}.

\subsubsection{\label{sec:Thermoremanent magnetization}Thermoremanent magnetization}

\begin{figure}
    \includegraphics[width=0.45\textwidth]{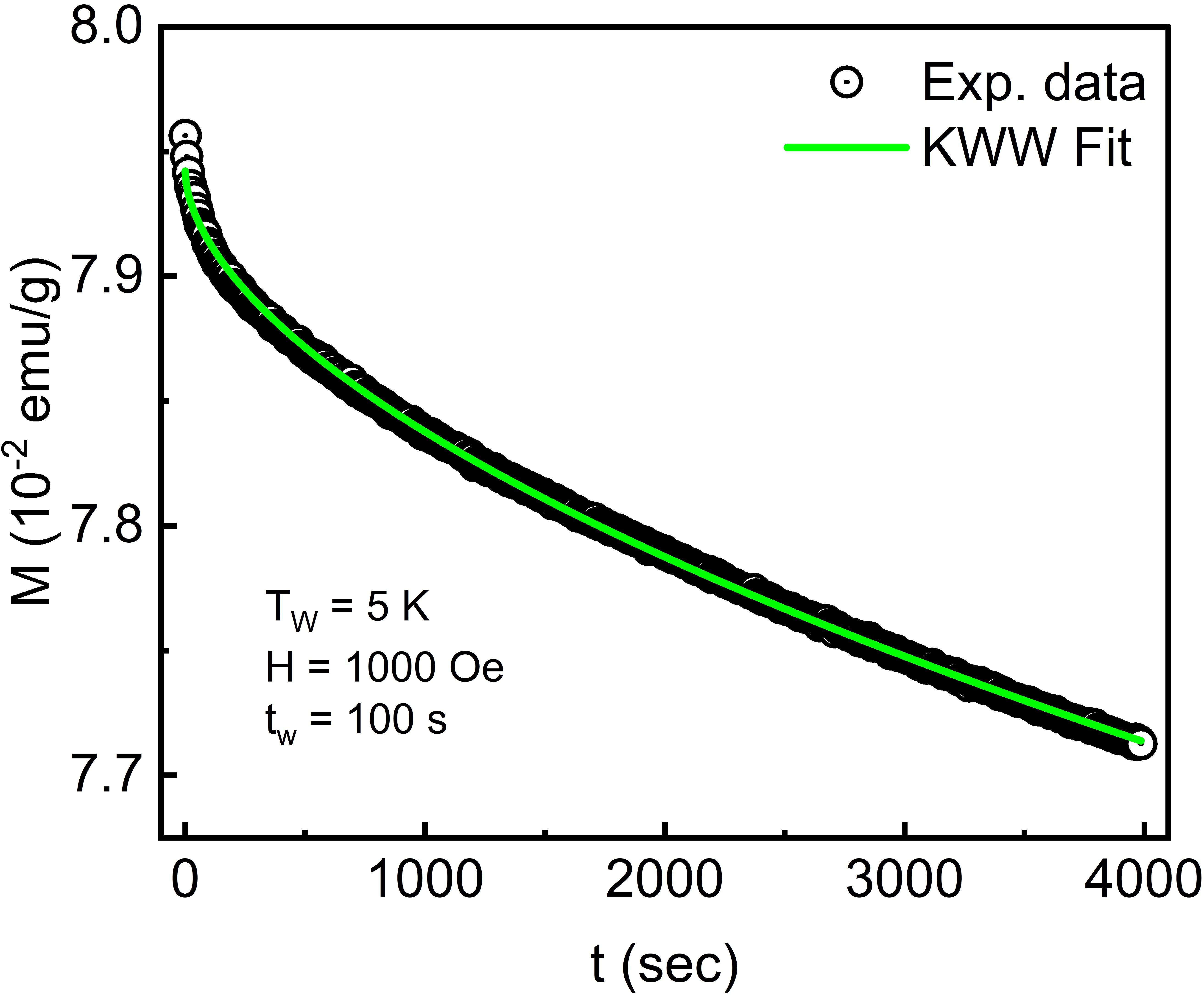}
    \caption{\label{fig:TRM}Relaxation of the field-cooled (FC) magnetization at 5 K after a holding time 100 s, as explained in the text. The green solid line reflects the fitting to the stretched exponential function, given by equation (\ref{eqn:7}).}
\end{figure}

To investigate the non-equilibrium spin dynamics in the alloy, thermoremanent magnetization (TRM) measurement is carried out in FC mode. The sample is cooled from 300 K to a holding temperature T$_W$ = 5 K below T$_P$ at a rate of 10 K/min. in an applied magnetic field of 1 kOe. After a holding time of 100 s, the field is rapidly turned off to zero, and the time-evolution of magnetization (M(t)) is measured. The measured magnetization M(t) at T$_W$ = 5 K after waiting for 100 s under FC mode is presented in Fig. \ref{fig:TRM}. To model the observed TRM, the Kohlrausch–Williams–Watt (KWW) equation \cite{Ref40, Ref41, Ref42, Ref43} is widely used, which is a well-known stretched exponential function applicable for SG systems.
\begin{equation}
  M(t)= M_0 + M_g ~exp⁡\left[-\left(\frac{t}{\tau}\right)^n \right],
  \label{eqn:7}
\end{equation}
where M$_0$, M$_g$, $\tau$ and n are the inherent magnetization, the parameter related to glassy component, characteristic relaxation time and stretched exponent (shape parameter), respectively. The energy barriers involved in the spin relaxation process are described by the value of n. The value of n lies typically between 0 and 1 for SGs \cite{Ref29, Ref40, Ref43, Ref44, Ref39, Ref43, Ref43a}. The green solid line in Fig. \ref{fig:TRM} depicts the least squares fit to the observed TRM data using equation (\ref{eqn:7}). The value of stretched exponent n for the studied Mn$_2$CuGe alloy is 0.56(6) that lies in the range of typical SGs reported in literature \cite{Ref18, Ref29, Ref37, Ref39, Ref40, Ref41, Ref42, Ref43, Ref43a, Ref44}.

\subsubsection{\label{sec:Magnetic memory and relaxation effect}Magnetic memory and relaxation effect}

\begin{figure}
    \includegraphics[width=0.45\textwidth]{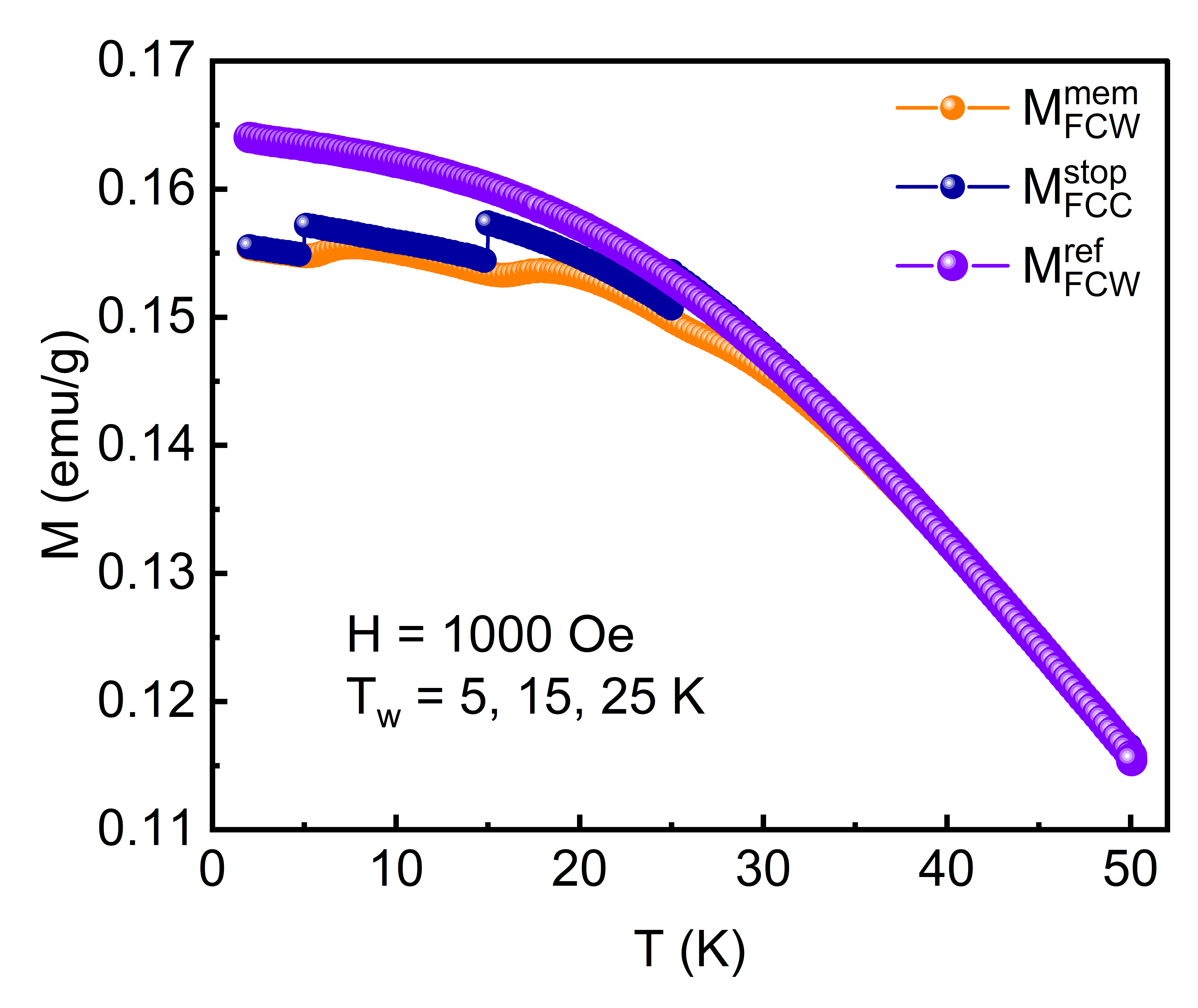}
    \caption{\label{fig:Memory}Magnetic memory effect of Mn$_2$CuGe alloy observed in FC mode at waiting temperatures 25 K, 15 K and 5 K for waiting time of 180 minutes. The memory curve clearly tries to trace the previous history of magnetization which is a typical signature of spin glass systems.}
\end{figure}

Memory effect measurement is performed in FC mode according to the protocol of Sun et. al \cite{Ref45} for the Mn$_2$CuGe alloy in order to gain insight into nonergodicity for SGs in the prepared alloy and to learn more about its low temperature dynamics. In the FC mode, the sample is cooled in the presence of applied magnetic field of 1 kOe from the paramagnetic state to the lowest temperature (2 K) with temporal stops at waiting temperatures T$_W$ = 25 K, 15 K and 5 K for waiting time t$_W$ = 180 minutes at each waiting temperature and the temperature dependent magnetization (M(T)) is recorded. On reaching the waiting temperatures, the magnetic field is switched off for the duration of waiting time (t$_W$ = 180 minutes). Further, the same magnetic field is re-applied and the cooling is resumed. This recorded magnetization is represented by M$_{FCC}^{stop}$ in Fig. \ref{fig:Memory} (solid sphere with line in blue color). After reaching 2 K, the sample is warmed to the paramagnetic region without any stops in the same magnetic field and the magnetization is recorded simultaneously, which is represented by M$_{FCW}^{mem}$ in Fig. \ref{fig:Memory} (solid sphere with line in orange color). For completeness, the magnetization M$_{FCW}^{ref}$ is also recorded (solid sphere with line in purple color). As observed from Fig. \ref{fig:Memory}, the magnetization M$_{FCW}^{mem}$ tries to follow  M$_{FCC}^{stop}$ curve and the M$_{FCW}^{mem}$ curve clearly portrays the previous history of magnetization. Thus, the Mn$_2$CuGe sample exhibits the FC magnetic memory effect below its SG freezing temperature which confirms its SG behavior. Similar FC magnetic memory effect is als reported in literature for other SG systems \cite{Ref37, Ref39, Ref41, Ref42, Ref46}.

\begin{figure*}
    \includegraphics[width=0.9\textwidth]{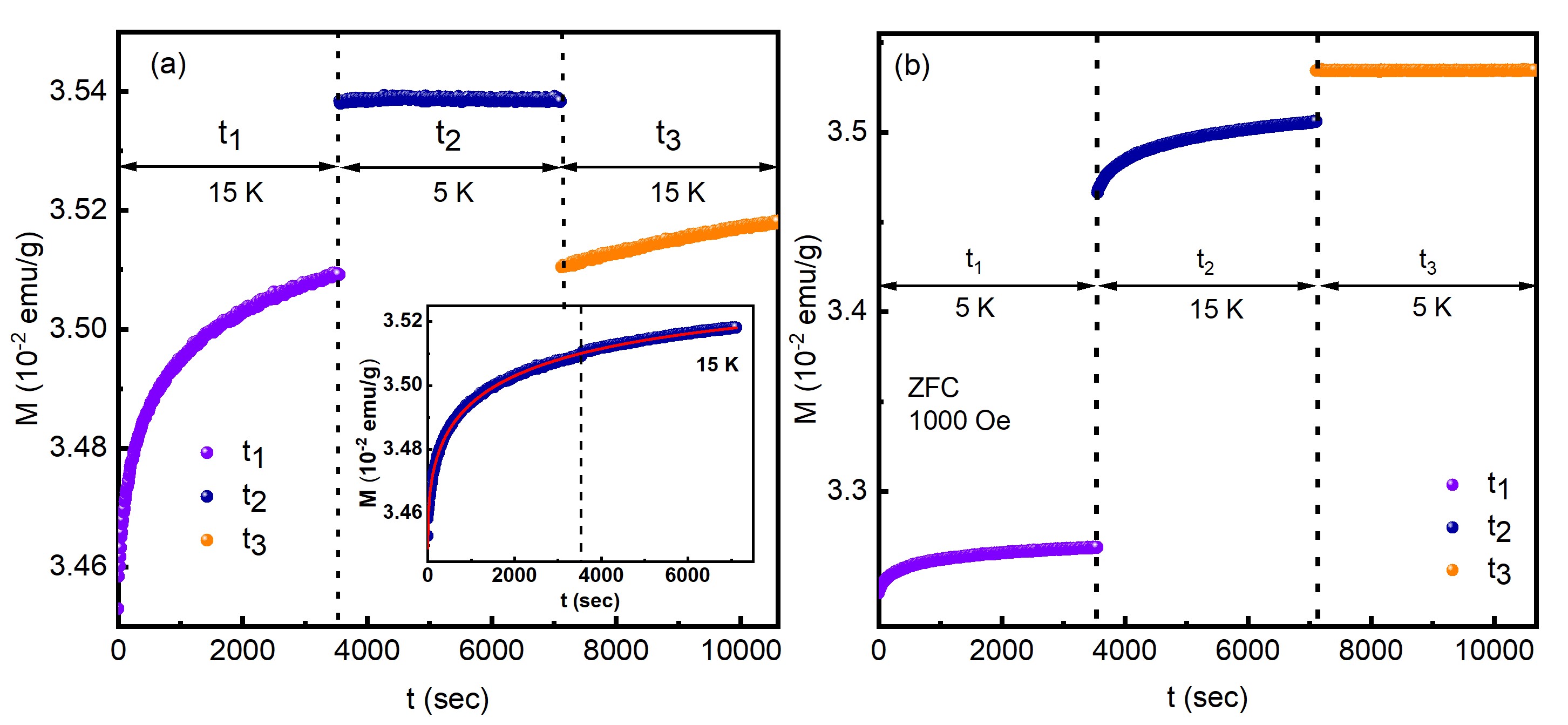}
    \caption{\label{fig:Relaxation}(a) Magnetic relaxation of Mn$_2$CuGe alloy at 15 K with intermediate temporary cooling to 5 K in 1 kOe applied magnetic field in the ZFC protocol. Inset represents the relaxation curve as a function of time at 15 K with fit to the stretched exponential function. The continuation of the relaxation behavior during t$_1$ is noticed during t$_3$ cycle. (b) Magnetic relaxation of Mn$_2$CuGe alloy at 5 K with intermediate temporary heating at 15 K in 1 kOe applied magnetic field in the ZFC mode.}
\end{figure*}

The nature of magnetic memory in SGs are widely explained using two models of spin relaxation namely, the hierarchical model and droplet model \cite{Ref39, Ref47, Ref48}. At a particular temperature, the hierarchical model suggests a multi-valley free energy landscape where the valleys merge upon warming while getting split into sub-valleys upon cooling. On the other hand, the droplet model predicts a two-valley structure with overall spin reversal on the free energy landscape at all temperatures below T$_P$. Experimentally, the hierarchical model shows asymmetric relaxation behavior, whereas droplet model reflects symmetric behavior for both increasing and decreasing temperature cycles. In order to understand these phenomena in Mn$_2$CuGe alloy, magnetic relaxation behavior in ZFC mode with intermediate cooling as well as intermediate heating is measured (Fig. \ref{fig:Relaxation}). In ZFC mode with intermediate cooling, the sample is first cooled to T$_1$ (15 K) $<$ T$_P$ in the absence of magnetic field. Once 15 K is reached, 1 kOe magnetic field is applied and magnetization is recorded as a function of time (M(t)) for t$_1$ = 60 minutes. After this step, the sample is quenched to a lower temperature T$_1$ - $\Delta$T (i.e. 5 K) in the same applied field (1 kOe) and the magnetization is again recorded as a function of time for a time period of t$_2$ = 60 minutes. Finally, the sample is reheated to T$_1$ = 15 K and the magnetization M(t) is recorded for t$_3$ = 60 minutes. The results of the above protocol are depicted in Fig. \ref{fig:Relaxation}\hyperref[fig:Relaxation]{(a)} and it can be seen that the magnetization value remains at the same level at the end of t$_1$ and at the start of t$_3$, which implies that the system remembers its magnetic state before and after intermediate cooling, as if there is no intermediate cooling performed. This type of magnetic relaxation behavior is reported earlier for many SG systems \cite{Ref18, Ref37, Ref39, Ref44}. As seen in the inset of Fig. \ref{fig:Relaxation}\hyperref[fig:Relaxation]{(a)}, the relaxation curve t$_3$ follows t$_1$ and the continued relaxation can be well fitted by equation (\ref{eqn:7}). The least square fitting (solid red line) to equation (\ref{eqn:7}) gives the value of the stretched exponent n = 0.4.  The value of n lies within the typical range for SG systems reported earlier \cite{Ref18, Ref29, Ref37, Ref40, Ref41, Ref42, Ref43, Ref44}.

Similar to the above protocol, the magnetic relaxation in ZFC mode with intermediate heating is also recorded. First, the sample is cooled to T$_1$ (5 K) $<$ T$_P$ in the absence of magnetic field. Once 5 K is reached, 1 kOe magnetic field is applied and magnetization is recorded as a function of time for t$_1$ = 60 minutes. After this step, the sample is heated to a higher temperature T$_1$ + $\Delta$T (i.e. 15 K) in the same applied field (1 kOe) and the magnetization is recorded as a function of time for the time period t$_2$ = 60 minutes. Finally, the sample is again cooled to T$_1$ = 5 K and the magnetization M(t) is recorded for t$_3$ = 60 minutes. The results of the above protocol are depicted in Fig. \ref{fig:Relaxation}\hyperref[fig:Relaxation]{(b)} and it can be seen that unlike the previous case, the magnetization value does not show a continuous behavior from t$_1$ to t$_3$ curve at 5 K, suggesting a different type of magnetic relaxation. This asymmetric response of temporary cooling and temporary heating confirms the hierarchical model of magnetic relaxation in Mn$_2$CuGe alloy. Similar type of magnetic relaxation is documented for other SGs \cite{Ref18, Ref37, Ref39, Ref49, Ref50}. The results of the magnetic relaxation and memory effect, therefore confirm the low-temperature SG nature in the studied Mn$_2$CuGe alloy.

\subsection{\label{sec:Temperature dependent specific heat}Temperature dependent specific heat}

\begin{figure}
    \includegraphics[width=0.45\textwidth]{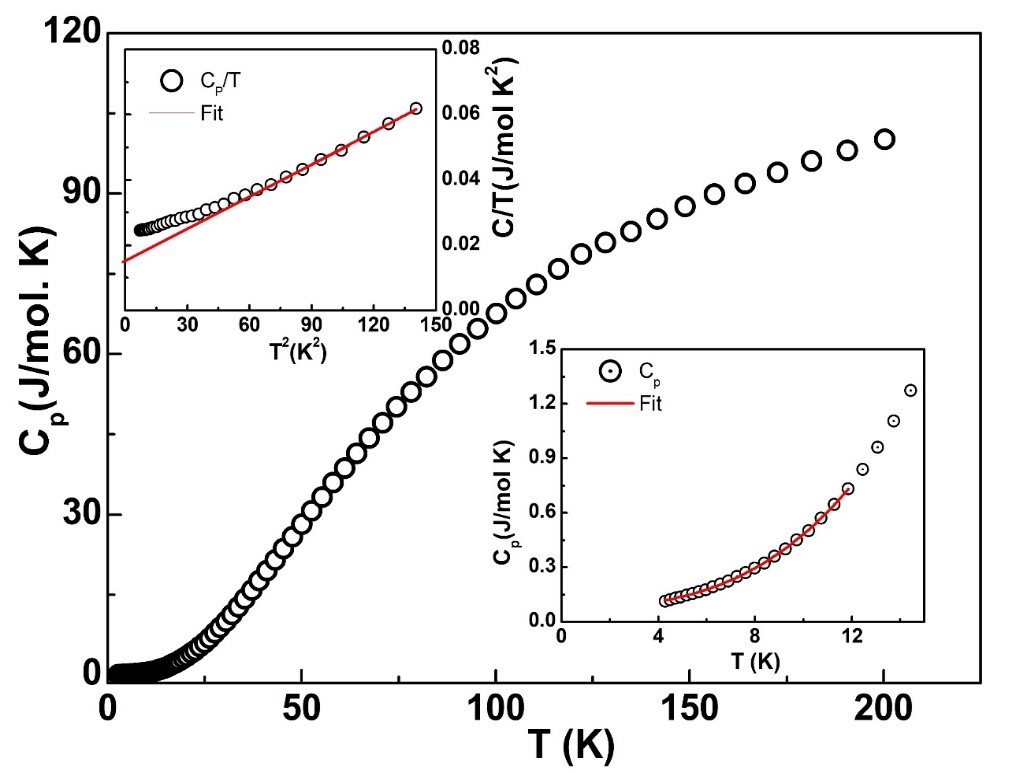}
    \caption{\label{fig:SH}Temperature dependent heat capacity for Mn$_2$CuGe. Upper Inset:  C$_p$/T versus T$^2$ and the red solid line is fit to C$_p$(T)/T = $\gamma$ + $\beta$T$^2$; Lower Inset: C$_p$ vs T and the red solid line is fit to C$_p$ = $\gamma$T + $\beta$T$^3$ + $\delta$T$^{3/2}$.}
\end{figure}
The analysis of low-temperature heat capacity data can provide further evidence of SG nature/magnetic long-range order. Therefore, heat capacity measurements on Mn$_2$CuGe sample are carried out at constant pressure in the absence of magnetic field in the temperature range of 1.8 K to 200 K. The temperature-dependent specific heat capacity of Mn$_2$CuGe is presented in Fig. \ref{fig:SH}. No anomaly associated with the magnetic long-range order is observed down to 2 K. The value of C$_p$ at T = 200 K is $\sim$ 100 J mol$^{-1}$ K$^{-1}$, which is close to the expected Dulong-Petit value C$_p$ = 3mR = 12R = 99.76 J mol$^{-1}$ K$^{-1}$, where R is the gas constant and m is the number of atoms per formula unit. The low-temperature heat capacity data is well described by C$_p$(T) = $\gamma$T + $\beta$T$^3$, where $\gamma$ is the Sommerfeld coefficient and $\beta$ is the Debye coefficient. In the above relation, $\gamma$T describes the electronic contribution to the heat capacity and $\beta$T$^3$ denotes the phonon contribution. The low-temperature C$_p$/T versus T$^2$ plot shown in the upper inset of Fig. \ref{fig:SH}, clearly depicts a nonlinear behavior. However, at very low temperatures, the C$_p$ data is fitted well by introducing a magnetic term $\delta$T$^{3/2}$ in C$_p$(T) = $\gamma$T + $\beta$T$^3$, thus modifying the relation as C$_p$ = $\gamma$T + $\beta$T$^3$ + $\delta$T$^{3/2}$, where $\delta$ is the coefficient of T$^{3/2}$ \cite{Ref51}. The term T$^{3/2}$ in C$_p$ is typical for SG and FM systems \cite{Ref52}. The lower inset of Fig. \ref{fig:SH} represents the best fit of the data in the temperature range of 2 K – 10 K which yields $\gamma$ = 78.3 mJ mol$^{-1}$ K$^{-2}$, $\beta$ = 0.81 mJ mol$^{-1}$ K$^{-4}$, and $\delta$ = 0.64 mJ mol$^{-1}$ K$^{-5/2}$. Using the value of $\beta$, the Debye temperature ($\theta_D$) can be evaluated by substituting the value of $\beta$ in $\theta_D$ = $(\frac{12\pi^4 Rm}{5\beta})^{1/3}$, where m is the number of atoms per formula unit (m = 4 in this case). The obtained value of $\theta_D$ is 212.38 K. The density of degenerate conduction carrier states at the Fermi level D(E$_F$) for both directions is obtained using the following relation \cite{Ref53, Ref54, Ref55}:
\begin{equation}
   D(E_F)=3\gamma/(\pi^2 k_{B}^{2}),
   \label{eqn:8}
\end{equation}                                         
which can be written as,
\begin{equation}
    D(E_F)[states/(eV f.u.)]=\frac{1}{2.359} \gamma[mJ/(mol.K^2)].
    \label{eqn:9}
\end{equation}

The density of states D(E$_F$) for Mn$_2$CuGe alloy is calculated from the $\gamma$ value using equation (\ref{eqn:9}) and is found to be 33.19 states/(eV.f.u). The large value of $\gamma$ in the electronic contribution to the specific heat indicates strong spin fluctuations due to the competing magnetic interactions. Such large values of $\gamma$ are reported for other Mn-based HAs such as Mn$_3$Al (69.4 mJ mol$^{-1}$ K$^{-2}$) \cite{Ref2c} and Mn$_2$FeAl (210 mJ mol$^{-1}$ K$^{-2}$) \cite{Ref56}. Furthermore, the obtained value of $\theta_D$ is smaller than those of Mn$_3$Al (240 K) \cite{Ref2c} and Mn$_2$FeAl (420 K) \cite{Ref56}.

\section{Conclusions}
In summary, a detailed investigation of the structural, magnetic and specific heat properties of arc-melted polycrystalline Mn$_2$CuGe Heusler alloy is presented through DC magnetization and heat capacity measurements. The structural analysis of the alloy through Rietveld refinement of room-temperature powder XRD data reveals a mixed hexagonal P3c1 (major) and P63/mmc (minor) phases. The analysis of DC magnetization at high temperatures suggests a PM - FiM transition at T$_C$ $\approx$ 682 K that is well supported by the N\'{e}el theory of ferrimagnets. The compensation of magnetization is achieved at T$_{COM} \approx$ 250 K. The existence of a magnetically frustrated SG character below the SG freezing temperature T$_P \approx$ 25.6 K is also confirmed by the analysis of DC magnetization such as the bifurcation of the ZFC and FC curves and the shifting of peak temperatures towards lower temperature with increasing applied magnetic field. The data points fall on the de Almeida-Thouless line in T$_P$ vs H$^{2/3}$ curve as expected for SG systems. The observed non-zero coercivity and remanence with narrow hysteresis in the isothermal M-H curves indicate competing FM - AFM interactions and frustrated magnetic ground state which, again supports the presence of low-temperature SG behavior in the alloy. A clear signature of slow relaxation and the magnetic memory effect in FC mode below the SG freezing temperature further confirms the low-temperature SG nature of the prepared alloy. The experimentally observed asymmetric response of the material's magnetic relaxation behavior below T$_P$ during intermediate heating and intermediate cooling cycles as a function of time in the ZFC mode favors the hierarchical model of spin relaxation in SG systems. The low-temperature SG character is further confirmed by the temperature-dependent heat capacity measurements which indicates a strong electronic contribution to specific heat which is probably due to the strong spin fluctuations resulting from competing magnetic interactions. Nevertheless, Mn$_2$CuGe being a high-temperature ferrimagnetic material, holds tremendous potential for futuristic spintronics-based applications. The current work is surely expected to encourage the experimental investigations of other theoretically predicted ferrimagnetic materials along with the designing and development of novel functional materials for technological innovations.

\begin{acknowledgments}
AKP acknowledges the support of SERB-DST, New Delhi, India (Grant no. CRG/2020/003590) and DST (Grant no. INT/RUS/RFBR/379). AKK acknowledges UGC New Delhi, India, for providing the financial support through a JRF Fellowship (16-6(DEC. 2018)/2019 NET/CSIR). Authors acknowledge the “Low temperature and high magnetic field facility under CIF” at the Central University of Rajasthan as well as the central instrumentation facility at IISER Bhopal for magnetic measurements.
\end{acknowledgments}

\bibliography{references}

\begin{thebibliography}{61}%
\makeatletter
\providecommand \@ifxundefined [1]{%
 \@ifx{#1\undefined}
}%
\providecommand \@ifnum [1]{%
 \ifnum #1\expandafter \@firstoftwo
 \else \expandafter \@secondoftwo
 \fi
}%
\providecommand \@ifx [1]{%
 \ifx #1\expandafter \@firstoftwo
 \else \expandafter \@secondoftwo
 \fi
}%
\providecommand \natexlab [1]{#1}%
\providecommand \enquote  [1]{``#1''}%
\providecommand \bibnamefont  [1]{#1}%
\providecommand \bibfnamefont [1]{#1}%
\providecommand \citenamefont [1]{#1}%
\providecommand \href@noop [0]{\@secondoftwo}%
\providecommand \href [0]{\begingroup \@sanitize@url \@href}%
\providecommand \@href[1]{\@@startlink{#1}\@@href}%
\providecommand \@@href[1]{\endgroup#1\@@endlink}%
\providecommand \@sanitize@url [0]{\catcode `\\12\catcode `\$12\catcode `\&12\catcode `\#12\catcode `\^12\catcode `\_12\catcode `\%12\relax}%
\providecommand \@@startlink[1]{}%
\providecommand \@@endlink[0]{}%
\providecommand \url  [0]{\begingroup\@sanitize@url \@url }%
\providecommand \@url [1]{\endgroup\@href {#1}{\urlprefix }}%
\providecommand \urlprefix  [0]{URL }%
\providecommand \Eprint [0]{\href }%
\providecommand \doibase [0]{https://doi.org/}%
\providecommand \selectlanguage [0]{\@gobble}%
\providecommand \bibinfo  [0]{\@secondoftwo}%
\providecommand \bibfield  [0]{\@secondoftwo}%
\providecommand \translation [1]{[#1]}%
\providecommand \BibitemOpen [0]{}%
\providecommand \bibitemStop [0]{}%
\providecommand \bibitemNoStop [0]{.\EOS\space}%
\providecommand \EOS [0]{\spacefactor3000\relax}%
\providecommand \BibitemShut  [1]{\csname bibitem#1\endcsname}%
\let\auto@bib@innerbib\@empty
%</preamble>
\bibitem [{\citenamefont {{\v{Z}}uti{\'c}}\ \emph {et~al.}(2004)\citenamefont {{\v{Z}}uti{\'c}}, \citenamefont {Fabian},\ and\ \citenamefont {Sarma}}]{Ref1}%
  \BibitemOpen
  \bibfield  {author} {\bibinfo {author} {\bibfnamefont {I.}~\bibnamefont {{\v{Z}}uti{\'c}}}, \bibinfo {author} {\bibfnamefont {J.}~\bibnamefont {Fabian}},\ and\ \bibinfo {author} {\bibfnamefont {S.~D.}\ \bibnamefont {Sarma}},\ }\href@noop {} {\bibfield  {journal} {\bibinfo  {journal} {Reviews of Modern Physics}\ }\textbf {\bibinfo {volume} {76}},\ \bibinfo {pages} {323} (\bibinfo {year} {2004})}\BibitemShut {NoStop}%
\bibitem [{\citenamefont {Kim}\ \emph {et~al.}(2022)\citenamefont {Kim}, \citenamefont {Beach}, \citenamefont {Lee}, \citenamefont {Ono}, \citenamefont {Rasing},\ and\ \citenamefont {Yang}}]{Ref2}%
  \BibitemOpen
  \bibfield  {author} {\bibinfo {author} {\bibfnamefont {S.~K.}\ \bibnamefont {Kim}}, \bibinfo {author} {\bibfnamefont {G.~S.}\ \bibnamefont {Beach}}, \bibinfo {author} {\bibfnamefont {K.-J.}\ \bibnamefont {Lee}}, \bibinfo {author} {\bibfnamefont {T.}~\bibnamefont {Ono}}, \bibinfo {author} {\bibfnamefont {T.}~\bibnamefont {Rasing}},\ and\ \bibinfo {author} {\bibfnamefont {H.}~\bibnamefont {Yang}},\ }\href@noop {} {\bibfield  {journal} {\bibinfo  {journal} {Nature Materials}\ }\textbf {\bibinfo {volume} {21}},\ \bibinfo {pages} {24} (\bibinfo {year} {2022})}\BibitemShut {NoStop}%
\bibitem [{\citenamefont {Jungwirth}\ \emph {et~al.}(2016)\citenamefont {Jungwirth}, \citenamefont {Marti}, \citenamefont {Wadley},\ and\ \citenamefont {Wunderlich}}]{Ref2a}%
  \BibitemOpen
  \bibfield  {author} {\bibinfo {author} {\bibfnamefont {T.}~\bibnamefont {Jungwirth}}, \bibinfo {author} {\bibfnamefont {X.}~\bibnamefont {Marti}}, \bibinfo {author} {\bibfnamefont {P.}~\bibnamefont {Wadley}},\ and\ \bibinfo {author} {\bibfnamefont {J.}~\bibnamefont {Wunderlich}},\ }\href@noop {} {\bibfield  {journal} {\bibinfo  {journal} {Nature Nanotechnology}\ }\textbf {\bibinfo {volume} {11}},\ \bibinfo {pages} {231} (\bibinfo {year} {2016})}\BibitemShut {NoStop}%
\bibitem [{\citenamefont {Baltz}\ \emph {et~al.}(2018)\citenamefont {Baltz}, \citenamefont {Manchon}, \citenamefont {Tsoi}, \citenamefont {Moriyama}, \citenamefont {Ono},\ and\ \citenamefont {Tserkovnyak}}]{Ref2b}%
  \BibitemOpen
  \bibfield  {author} {\bibinfo {author} {\bibfnamefont {V.}~\bibnamefont {Baltz}}, \bibinfo {author} {\bibfnamefont {A.}~\bibnamefont {Manchon}}, \bibinfo {author} {\bibfnamefont {M.}~\bibnamefont {Tsoi}}, \bibinfo {author} {\bibfnamefont {T.}~\bibnamefont {Moriyama}}, \bibinfo {author} {\bibfnamefont {T.}~\bibnamefont {Ono}},\ and\ \bibinfo {author} {\bibfnamefont {Y.}~\bibnamefont {Tserkovnyak}},\ }\href@noop {} {\bibfield  {journal} {\bibinfo  {journal} {Reviews of Modern Physics}\ }\textbf {\bibinfo {volume} {90}},\ \bibinfo {pages} {015005} (\bibinfo {year} {2018})}\BibitemShut {NoStop}%
\bibitem [{\citenamefont {Dash}\ \emph {et~al.}(2024)\citenamefont {Dash}, \citenamefont {Vasundhara},\ and\ \citenamefont {Patra}}]{Ref2c}%
  \BibitemOpen
  \bibfield  {author} {\bibinfo {author} {\bibfnamefont {S.}~\bibnamefont {Dash}}, \bibinfo {author} {\bibfnamefont {M.}~\bibnamefont {Vasundhara}},\ and\ \bibinfo {author} {\bibfnamefont {A.~K.}\ \bibnamefont {Patra}},\ }\href@noop {} {\bibfield  {journal} {\bibinfo  {journal} {Journal of Magnetism and Magnetic Materials}\ }\textbf {\bibinfo {volume} {589}},\ \bibinfo {pages} {171474} (\bibinfo {year} {2024})}\BibitemShut {NoStop}%
\bibitem [{\citenamefont {Venkateswara}\ \emph {et~al.}(2018)\citenamefont {Venkateswara}, \citenamefont {Gupta}, \citenamefont {Samatham}, \citenamefont {Varma}, \citenamefont {Enamullah}, \citenamefont {Suresh},\ and\ \citenamefont {Alam}}]{Ref3}%
  \BibitemOpen
  \bibfield  {author} {\bibinfo {author} {\bibfnamefont {Y.}~\bibnamefont {Venkateswara}}, \bibinfo {author} {\bibfnamefont {S.}~\bibnamefont {Gupta}}, \bibinfo {author} {\bibfnamefont {S.~S.}\ \bibnamefont {Samatham}}, \bibinfo {author} {\bibfnamefont {M.~R.}\ \bibnamefont {Varma}}, \bibinfo {author} {\bibnamefont {Enamullah}}, \bibinfo {author} {\bibfnamefont {K.}~\bibnamefont {Suresh}},\ and\ \bibinfo {author} {\bibfnamefont {A.}~\bibnamefont {Alam}},\ }\href@noop {} {\bibfield  {journal} {\bibinfo  {journal} {Physical Review B}\ }\textbf {\bibinfo {volume} {97}},\ \bibinfo {pages} {054407} (\bibinfo {year} {2018})}\BibitemShut {NoStop}%
\bibitem [{\citenamefont {Graf}\ \emph {et~al.}(2010)\citenamefont {Graf}, \citenamefont {Parkin},\ and\ \citenamefont {Felser}}]{Ref4}%
  \BibitemOpen
  \bibfield  {author} {\bibinfo {author} {\bibfnamefont {T.}~\bibnamefont {Graf}}, \bibinfo {author} {\bibfnamefont {S.~S.}\ \bibnamefont {Parkin}},\ and\ \bibinfo {author} {\bibfnamefont {C.}~\bibnamefont {Felser}},\ }\href@noop {} {\bibfield  {journal} {\bibinfo  {journal} {IEEE Transactions on Magnetics}\ }\textbf {\bibinfo {volume} {47}},\ \bibinfo {pages} {367} (\bibinfo {year} {2010})}\BibitemShut {NoStop}%
\bibitem [{\citenamefont {Wollmann}\ \emph {et~al.}(2017)\citenamefont {Wollmann}, \citenamefont {Nayak}, \citenamefont {Parkin},\ and\ \citenamefont {Felser}}]{Ref5}%
  \BibitemOpen
  \bibfield  {author} {\bibinfo {author} {\bibfnamefont {L.}~\bibnamefont {Wollmann}}, \bibinfo {author} {\bibfnamefont {A.~K.}\ \bibnamefont {Nayak}}, \bibinfo {author} {\bibfnamefont {S.~S.}\ \bibnamefont {Parkin}},\ and\ \bibinfo {author} {\bibfnamefont {C.}~\bibnamefont {Felser}},\ }\href@noop {} {\bibfield  {journal} {\bibinfo  {journal} {Annual Review of Materials Research}\ }\textbf {\bibinfo {volume} {47}},\ \bibinfo {pages} {247} (\bibinfo {year} {2017})}\BibitemShut {NoStop}%
\bibitem [{\citenamefont {Manna}\ \emph {et~al.}(2018)\citenamefont {Manna}, \citenamefont {Sun}, \citenamefont {Muechler}, \citenamefont {K{\"u}bler},\ and\ \citenamefont {Felser}}]{Ref5a}%
  \BibitemOpen
  \bibfield  {author} {\bibinfo {author} {\bibfnamefont {K.}~\bibnamefont {Manna}}, \bibinfo {author} {\bibfnamefont {Y.}~\bibnamefont {Sun}}, \bibinfo {author} {\bibfnamefont {L.}~\bibnamefont {Muechler}}, \bibinfo {author} {\bibfnamefont {J.}~\bibnamefont {K{\"u}bler}},\ and\ \bibinfo {author} {\bibfnamefont {C.}~\bibnamefont {Felser}},\ }\href@noop {} {\bibfield  {journal} {\bibinfo  {journal} {Nature Reviews Materials}\ }\textbf {\bibinfo {volume} {3}},\ \bibinfo {pages} {244} (\bibinfo {year} {2018})}\BibitemShut {NoStop}%
\bibitem [{\citenamefont {Weht}\ and\ \citenamefont {Pickett}(1999)}]{Ref6}%
  \BibitemOpen
  \bibfield  {author} {\bibinfo {author} {\bibfnamefont {R.}~\bibnamefont {Weht}}\ and\ \bibinfo {author} {\bibfnamefont {W.~E.}\ \bibnamefont {Pickett}},\ }\href@noop {} {\bibfield  {journal} {\bibinfo  {journal} {Physical Review B}\ }\textbf {\bibinfo {volume} {60}},\ \bibinfo {pages} {13006} (\bibinfo {year} {1999})}\BibitemShut {NoStop}%
\bibitem [{\citenamefont {{\"O}zdogan}\ \emph {et~al.}(2006)\citenamefont {{\"O}zdogan}, \citenamefont {Galanakis}, \citenamefont {{\c{S}}a{\c{s}}ioglu},\ and\ \citenamefont {Akta{\c{s}}}}]{Ref7}%
  \BibitemOpen
  \bibfield  {author} {\bibinfo {author} {\bibfnamefont {K.}~\bibnamefont {{\"O}zdogan}}, \bibinfo {author} {\bibfnamefont {I.}~\bibnamefont {Galanakis}}, \bibinfo {author} {\bibfnamefont {E.}~\bibnamefont {{\c{S}}a{\c{s}}ioglu}},\ and\ \bibinfo {author} {\bibfnamefont {B.}~\bibnamefont {Akta{\c{s}}}},\ }\href@noop {} {\bibfield  {journal} {\bibinfo  {journal} {Journal of Physics: Condensed Matter}\ }\textbf {\bibinfo {volume} {18}},\ \bibinfo {pages} {2905} (\bibinfo {year} {2006})}\BibitemShut {NoStop}%
\bibitem [{\citenamefont {Luo}\ \emph {et~al.}(2008{\natexlab{a}})\citenamefont {Luo}, \citenamefont {Zhu}, \citenamefont {Liu}, \citenamefont {Xu}, \citenamefont {Wu}, \citenamefont {Liu}, \citenamefont {Qu},\ and\ \citenamefont {Li}}]{Ref8}%
  \BibitemOpen
  \bibfield  {author} {\bibinfo {author} {\bibfnamefont {H.}~\bibnamefont {Luo}}, \bibinfo {author} {\bibfnamefont {Z.}~\bibnamefont {Zhu}}, \bibinfo {author} {\bibfnamefont {G.}~\bibnamefont {Liu}}, \bibinfo {author} {\bibfnamefont {S.}~\bibnamefont {Xu}}, \bibinfo {author} {\bibfnamefont {G.}~\bibnamefont {Wu}}, \bibinfo {author} {\bibfnamefont {H.}~\bibnamefont {Liu}}, \bibinfo {author} {\bibfnamefont {J.}~\bibnamefont {Qu}},\ and\ \bibinfo {author} {\bibfnamefont {Y.}~\bibnamefont {Li}},\ }\href@noop {} {\bibfield  {journal} {\bibinfo  {journal} {Journal of Magnetism and Magnetic Materials}\ }\textbf {\bibinfo {volume} {320}},\ \bibinfo {pages} {421} (\bibinfo {year} {2008}{\natexlab{a}})}\BibitemShut {NoStop}%
\bibitem [{\citenamefont {Luo}\ \emph {et~al.}(2008{\natexlab{b}})\citenamefont {Luo}, \citenamefont {Zhang}, \citenamefont {Zhu}, \citenamefont {Ma}, \citenamefont {Xu}, \citenamefont {Wu}, \citenamefont {Zhu}, \citenamefont {Jiang},\ and\ \citenamefont {Xu}}]{Ref9}%
  \BibitemOpen
  \bibfield  {author} {\bibinfo {author} {\bibfnamefont {H.}~\bibnamefont {Luo}}, \bibinfo {author} {\bibfnamefont {H.}~\bibnamefont {Zhang}}, \bibinfo {author} {\bibfnamefont {Z.}~\bibnamefont {Zhu}}, \bibinfo {author} {\bibfnamefont {L.}~\bibnamefont {Ma}}, \bibinfo {author} {\bibfnamefont {S.}~\bibnamefont {Xu}}, \bibinfo {author} {\bibfnamefont {G.}~\bibnamefont {Wu}}, \bibinfo {author} {\bibfnamefont {X.}~\bibnamefont {Zhu}}, \bibinfo {author} {\bibfnamefont {C.}~\bibnamefont {Jiang}},\ and\ \bibinfo {author} {\bibfnamefont {H.}~\bibnamefont {Xu}},\ }\href@noop {} {\bibfield  {journal} {\bibinfo  {journal} {Journal of Applied Physics}\ }\textbf {\bibinfo {volume} {103}} (\bibinfo {year} {2008}{\natexlab{b}})}\BibitemShut {NoStop}%
\bibitem [{\citenamefont {Xing}\ \emph {et~al.}(2008)\citenamefont {Xing}, \citenamefont {Li}, \citenamefont {Dong}, \citenamefont {Long},\ and\ \citenamefont {Zhang}}]{Ref10}%
  \BibitemOpen
  \bibfield  {author} {\bibinfo {author} {\bibfnamefont {N.}~\bibnamefont {Xing}}, \bibinfo {author} {\bibfnamefont {H.}~\bibnamefont {Li}}, \bibinfo {author} {\bibfnamefont {J.}~\bibnamefont {Dong}}, \bibinfo {author} {\bibfnamefont {R.}~\bibnamefont {Long}},\ and\ \bibinfo {author} {\bibfnamefont {C.}~\bibnamefont {Zhang}},\ }\href@noop {} {\bibfield  {journal} {\bibinfo  {journal} {Computational Materials Science}\ }\textbf {\bibinfo {volume} {42}},\ \bibinfo {pages} {600} (\bibinfo {year} {2008})}\BibitemShut {NoStop}%
\bibitem [{\citenamefont {Luo}\ \emph {et~al.}(2008{\natexlab{c}})\citenamefont {Luo}, \citenamefont {Zhu}, \citenamefont {Ma}, \citenamefont {Xu}, \citenamefont {Zhu}, \citenamefont {Jiang}, \citenamefont {Xu},\ and\ \citenamefont {Wu}}]{Ref11}%
  \BibitemOpen
  \bibfield  {author} {\bibinfo {author} {\bibfnamefont {H.}~\bibnamefont {Luo}}, \bibinfo {author} {\bibfnamefont {Z.}~\bibnamefont {Zhu}}, \bibinfo {author} {\bibfnamefont {L.}~\bibnamefont {Ma}}, \bibinfo {author} {\bibfnamefont {S.}~\bibnamefont {Xu}}, \bibinfo {author} {\bibfnamefont {X.}~\bibnamefont {Zhu}}, \bibinfo {author} {\bibfnamefont {C.}~\bibnamefont {Jiang}}, \bibinfo {author} {\bibfnamefont {H.}~\bibnamefont {Xu}},\ and\ \bibinfo {author} {\bibfnamefont {G.}~\bibnamefont {Wu}},\ }\href@noop {} {\bibfield  {journal} {\bibinfo  {journal} {Journal of Physics D: Applied Physics}\ }\textbf {\bibinfo {volume} {41}},\ \bibinfo {pages} {055010} (\bibinfo {year} {2008}{\natexlab{c}})}\BibitemShut {NoStop}%
\bibitem [{\citenamefont {Dai}\ \emph {et~al.}(2006)\citenamefont {Dai}, \citenamefont {Liu}, \citenamefont {Chen}, \citenamefont {Chen},\ and\ \citenamefont {Wu}}]{Ref12}%
  \BibitemOpen
  \bibfield  {author} {\bibinfo {author} {\bibfnamefont {X.}~\bibnamefont {Dai}}, \bibinfo {author} {\bibfnamefont {G.}~\bibnamefont {Liu}}, \bibinfo {author} {\bibfnamefont {L.}~\bibnamefont {Chen}}, \bibinfo {author} {\bibfnamefont {J.}~\bibnamefont {Chen}},\ and\ \bibinfo {author} {\bibfnamefont {G.}~\bibnamefont {Wu}},\ }\href@noop {} {\bibfield  {journal} {\bibinfo  {journal} {Solid state communications}\ }\textbf {\bibinfo {volume} {140}},\ \bibinfo {pages} {533} (\bibinfo {year} {2006})}\BibitemShut {NoStop}%
\bibitem [{\citenamefont {Liu}\ \emph {et~al.}(2008)\citenamefont {Liu}, \citenamefont {Dai}, \citenamefont {Liu}, \citenamefont {Chen}, \citenamefont {Li}, \citenamefont {Xiao},\ and\ \citenamefont {Wu}}]{Ref13}%
  \BibitemOpen
  \bibfield  {author} {\bibinfo {author} {\bibfnamefont {G.}~\bibnamefont {Liu}}, \bibinfo {author} {\bibfnamefont {X.}~\bibnamefont {Dai}}, \bibinfo {author} {\bibfnamefont {H.}~\bibnamefont {Liu}}, \bibinfo {author} {\bibfnamefont {J.}~\bibnamefont {Chen}}, \bibinfo {author} {\bibfnamefont {Y.}~\bibnamefont {Li}}, \bibinfo {author} {\bibfnamefont {G.}~\bibnamefont {Xiao}},\ and\ \bibinfo {author} {\bibfnamefont {G.}~\bibnamefont {Wu}},\ }\href@noop {} {\bibfield  {journal} {\bibinfo  {journal} {Physical Review B}\ }\textbf {\bibinfo {volume} {77}},\ \bibinfo {pages} {014424} (\bibinfo {year} {2008})}\BibitemShut {NoStop}%
\bibitem [{\citenamefont {Mouchou}\ \emph {et~al.}(2024)\citenamefont {Mouchou}, \citenamefont {Toual}, \citenamefont {Khatiri}, \citenamefont {Azouaoui}, \citenamefont {Bouslykhane}, \citenamefont {Hourmatallah},\ and\ \citenamefont {Benzakour}}]{Ref14}%
  \BibitemOpen
  \bibfield  {author} {\bibinfo {author} {\bibfnamefont {S.}~\bibnamefont {Mouchou}}, \bibinfo {author} {\bibfnamefont {Y.}~\bibnamefont {Toual}}, \bibinfo {author} {\bibfnamefont {M.}~\bibnamefont {Khatiri}}, \bibinfo {author} {\bibfnamefont {A.}~\bibnamefont {Azouaoui}}, \bibinfo {author} {\bibfnamefont {K.}~\bibnamefont {Bouslykhane}}, \bibinfo {author} {\bibfnamefont {A.}~\bibnamefont {Hourmatallah}},\ and\ \bibinfo {author} {\bibfnamefont {N.}~\bibnamefont {Benzakour}},\ }\href@noop {} {\bibfield  {journal} {\bibinfo  {journal} {Materials Chemistry and Physics}\ }\textbf {\bibinfo {volume} {320}},\ \bibinfo {pages} {129455} (\bibinfo {year} {2024})}\BibitemShut {NoStop}%
\bibitem [{\citenamefont {Muthui}\ \emph {et~al.}(2017)\citenamefont {Muthui}, \citenamefont {Musembi}, \citenamefont {Mwabora}, \citenamefont {Skomski},\ and\ \citenamefont {Kashyap}}]{Ref15}%
  \BibitemOpen
  \bibfield  {author} {\bibinfo {author} {\bibfnamefont {Z.~W.}\ \bibnamefont {Muthui}}, \bibinfo {author} {\bibfnamefont {R.~J.}\ \bibnamefont {Musembi}}, \bibinfo {author} {\bibfnamefont {J.~M.}\ \bibnamefont {Mwabora}}, \bibinfo {author} {\bibfnamefont {R.}~\bibnamefont {Skomski}},\ and\ \bibinfo {author} {\bibfnamefont {A.}~\bibnamefont {Kashyap}},\ }\href@noop {} {\bibfield  {journal} {\bibinfo  {journal} {IEEE Transactions on Magnetics}\ }\textbf {\bibinfo {volume} {54}},\ \bibinfo {pages} {1} (\bibinfo {year} {2017})}\BibitemShut {NoStop}%
\bibitem [{\citenamefont {Nayak}\ \emph {et~al.}(2013)\citenamefont {Nayak}, \citenamefont {Nicklas}, \citenamefont {Chadov}, \citenamefont {Shekhar}, \citenamefont {Skourski}, \citenamefont {Winterlik},\ and\ \citenamefont {Felser}}]{Ref16}%
  \BibitemOpen
  \bibfield  {author} {\bibinfo {author} {\bibfnamefont {A.~K.}\ \bibnamefont {Nayak}}, \bibinfo {author} {\bibfnamefont {M.}~\bibnamefont {Nicklas}}, \bibinfo {author} {\bibfnamefont {S.}~\bibnamefont {Chadov}}, \bibinfo {author} {\bibfnamefont {C.}~\bibnamefont {Shekhar}}, \bibinfo {author} {\bibfnamefont {Y.}~\bibnamefont {Skourski}}, \bibinfo {author} {\bibfnamefont {J.}~\bibnamefont {Winterlik}},\ and\ \bibinfo {author} {\bibfnamefont {C.}~\bibnamefont {Felser}},\ }\href@noop {} {\bibfield  {journal} {\bibinfo  {journal} {Physical Review Letters}\ }\textbf {\bibinfo {volume} {110}},\ \bibinfo {pages} {127204} (\bibinfo {year} {2013})}\BibitemShut {NoStop}%
\bibitem [{\citenamefont {Liu}\ \emph {et~al.}(2016)\citenamefont {Liu}, \citenamefont {Zhang}, \citenamefont {Zhang}, \citenamefont {Zhang},\ and\ \citenamefont {Ma}}]{Ref17}%
  \BibitemOpen
  \bibfield  {author} {\bibinfo {author} {\bibfnamefont {Z.}~\bibnamefont {Liu}}, \bibinfo {author} {\bibfnamefont {Y.}~\bibnamefont {Zhang}}, \bibinfo {author} {\bibfnamefont {H.}~\bibnamefont {Zhang}}, \bibinfo {author} {\bibfnamefont {X.}~\bibnamefont {Zhang}},\ and\ \bibinfo {author} {\bibfnamefont {X.}~\bibnamefont {Ma}},\ }\href@noop {} {\bibfield  {journal} {\bibinfo  {journal} {Applied Physics Letters}\ }\textbf {\bibinfo {volume} {109}} (\bibinfo {year} {2016})}\BibitemShut {NoStop}%
\bibitem [{\citenamefont {Nayak}\ \emph {et~al.}(2012)\citenamefont {Nayak}, \citenamefont {Shekhar}, \citenamefont {Winterlik}, \citenamefont {Gupta},\ and\ \citenamefont {Felser}}]{Ref18}%
  \BibitemOpen
  \bibfield  {author} {\bibinfo {author} {\bibfnamefont {A.~K.}\ \bibnamefont {Nayak}}, \bibinfo {author} {\bibfnamefont {C.}~\bibnamefont {Shekhar}}, \bibinfo {author} {\bibfnamefont {J.}~\bibnamefont {Winterlik}}, \bibinfo {author} {\bibfnamefont {A.}~\bibnamefont {Gupta}},\ and\ \bibinfo {author} {\bibfnamefont {C.}~\bibnamefont {Felser}},\ }\href@noop {} {\bibfield  {journal} {\bibinfo  {journal} {Applied Physics Letters}\ }\textbf {\bibinfo {volume} {100}} (\bibinfo {year} {2012})}\BibitemShut {NoStop}%
\bibitem [{\citenamefont {Patel}\ \emph {et~al.}(2022)\citenamefont {Patel}, \citenamefont {Samatham}, \citenamefont {Lukoyanov}, \citenamefont {Babu},\ and\ \citenamefont {Suresh}}]{Ref19}%
  \BibitemOpen
  \bibfield  {author} {\bibinfo {author} {\bibfnamefont {A.~K.}\ \bibnamefont {Patel}}, \bibinfo {author} {\bibfnamefont {S.~S.}\ \bibnamefont {Samatham}}, \bibinfo {author} {\bibfnamefont {A.~V.}\ \bibnamefont {Lukoyanov}}, \bibinfo {author} {\bibfnamefont {P.}~\bibnamefont {Babu}},\ and\ \bibinfo {author} {\bibfnamefont {K.}~\bibnamefont {Suresh}},\ }\href@noop {} {\bibfield  {journal} {\bibinfo  {journal} {Physical Chemistry Chemical Physics}\ }\textbf {\bibinfo {volume} {24}},\ \bibinfo {pages} {29539} (\bibinfo {year} {2022})}\BibitemShut {NoStop}%
\bibitem [{\citenamefont {Gavrea}\ \emph {et~al.}(2020)\citenamefont {Gavrea}, \citenamefont {Hirian}, \citenamefont {Isnard}, \citenamefont {Pop},\ and\ \citenamefont {Benea}}]{Ref20}%
  \BibitemOpen
  \bibfield  {author} {\bibinfo {author} {\bibfnamefont {R.}~\bibnamefont {Gavrea}}, \bibinfo {author} {\bibfnamefont {R.}~\bibnamefont {Hirian}}, \bibinfo {author} {\bibfnamefont {O.}~\bibnamefont {Isnard}}, \bibinfo {author} {\bibfnamefont {V.}~\bibnamefont {Pop}},\ and\ \bibinfo {author} {\bibfnamefont {D.}~\bibnamefont {Benea}},\ }\href@noop {} {\bibfield  {journal} {\bibinfo  {journal} {Solid State Communications}\ }\textbf {\bibinfo {volume} {309}},\ \bibinfo {pages} {113812} (\bibinfo {year} {2020})}\BibitemShut {NoStop}%
\bibitem [{\citenamefont {Midhunlal}\ \emph {et~al.}(2018)\citenamefont {Midhunlal}, \citenamefont {Chelvane}, \citenamefont {Prabhu}, \citenamefont {Gopalan} \emph {et~al.}}]{Ref21}%
  \BibitemOpen
  \bibfield  {author} {\bibinfo {author} {\bibfnamefont {P.}~\bibnamefont {Midhunlal}}, \bibinfo {author} {\bibfnamefont {J.~A.}\ \bibnamefont {Chelvane}}, \bibinfo {author} {\bibfnamefont {D.}~\bibnamefont {Prabhu}}, \bibinfo {author} {\bibfnamefont {R.}~\bibnamefont {Gopalan}}, \emph {et~al.},\ }\href@noop {} {\bibfield  {journal} {\bibinfo  {journal} {arXiv preprint arXiv:1812.00714}\ } (\bibinfo {year} {2018})}\BibitemShut {NoStop}%
\bibitem [{\citenamefont {Stinshoff}\ \emph {et~al.}(2017)\citenamefont {Stinshoff}, \citenamefont {Nayak}, \citenamefont {Fecher}, \citenamefont {Balke}, \citenamefont {Ouardi}, \citenamefont {Skourski}, \citenamefont {Nakamura},\ and\ \citenamefont {Felser}}]{Ref22}%
  \BibitemOpen
  \bibfield  {author} {\bibinfo {author} {\bibfnamefont {R.}~\bibnamefont {Stinshoff}}, \bibinfo {author} {\bibfnamefont {A.~K.}\ \bibnamefont {Nayak}}, \bibinfo {author} {\bibfnamefont {G.~H.}\ \bibnamefont {Fecher}}, \bibinfo {author} {\bibfnamefont {B.}~\bibnamefont {Balke}}, \bibinfo {author} {\bibfnamefont {S.}~\bibnamefont {Ouardi}}, \bibinfo {author} {\bibfnamefont {Y.}~\bibnamefont {Skourski}}, \bibinfo {author} {\bibfnamefont {T.}~\bibnamefont {Nakamura}},\ and\ \bibinfo {author} {\bibfnamefont {C.}~\bibnamefont {Felser}},\ }\href@noop {} {\bibfield  {journal} {\bibinfo  {journal} {Physical Review B}\ }\textbf {\bibinfo {volume} {95}},\ \bibinfo {pages} {060410} (\bibinfo {year} {2017})}\BibitemShut {NoStop}%
\bibitem [{\citenamefont {Luo}\ \emph {et~al.}(2011)\citenamefont {Luo}, \citenamefont {Liu}, \citenamefont {Meng}, \citenamefont {Wang}, \citenamefont {Liu}, \citenamefont {Wu}, \citenamefont {Zhu},\ and\ \citenamefont {Jiang}}]{Ref23}%
  \BibitemOpen
  \bibfield  {author} {\bibinfo {author} {\bibfnamefont {H.}~\bibnamefont {Luo}}, \bibinfo {author} {\bibfnamefont {G.}~\bibnamefont {Liu}}, \bibinfo {author} {\bibfnamefont {F.}~\bibnamefont {Meng}}, \bibinfo {author} {\bibfnamefont {L.}~\bibnamefont {Wang}}, \bibinfo {author} {\bibfnamefont {E.}~\bibnamefont {Liu}}, \bibinfo {author} {\bibfnamefont {G.}~\bibnamefont {Wu}}, \bibinfo {author} {\bibfnamefont {X.}~\bibnamefont {Zhu}},\ and\ \bibinfo {author} {\bibfnamefont {C.}~\bibnamefont {Jiang}},\ }\href@noop {} {\bibfield  {journal} {\bibinfo  {journal} {Computational Materials Science}\ }\textbf {\bibinfo {volume} {50}},\ \bibinfo {pages} {3119} (\bibinfo {year} {2011})}\BibitemShut {NoStop}%
\bibitem [{\citenamefont {Li}\ \emph {et~al.}(2009)\citenamefont {Li}, \citenamefont {Ren}, \citenamefont {Zhang},\ and\ \citenamefont {Cao}}]{Ref24}%
  \BibitemOpen
  \bibfield  {author} {\bibinfo {author} {\bibfnamefont {S.}~\bibnamefont {Li}}, \bibinfo {author} {\bibfnamefont {Z.}~\bibnamefont {Ren}}, \bibinfo {author} {\bibfnamefont {X.}~\bibnamefont {Zhang}},\ and\ \bibinfo {author} {\bibfnamefont {C.}~\bibnamefont {Cao}},\ }\href@noop {} {\bibfield  {journal} {\bibinfo  {journal} {Physica B: Condensed Matter}\ }\textbf {\bibinfo {volume} {404}},\ \bibinfo {pages} {1965} (\bibinfo {year} {2009})}\BibitemShut {NoStop}%
\bibitem [{\citenamefont {Feng}\ \emph {et~al.}(2010)\citenamefont {Feng}, \citenamefont {Luo}, \citenamefont {Wang}, \citenamefont {Li}, \citenamefont {Zhu}, \citenamefont {Wu},\ and\ \citenamefont {Meng}}]{Ref25}%
  \BibitemOpen
  \bibfield  {author} {\bibinfo {author} {\bibfnamefont {Z.}~\bibnamefont {Feng}}, \bibinfo {author} {\bibfnamefont {H.}~\bibnamefont {Luo}}, \bibinfo {author} {\bibfnamefont {Y.}~\bibnamefont {Wang}}, \bibinfo {author} {\bibfnamefont {Y.}~\bibnamefont {Li}}, \bibinfo {author} {\bibfnamefont {W.}~\bibnamefont {Zhu}}, \bibinfo {author} {\bibfnamefont {G.}~\bibnamefont {Wu}},\ and\ \bibinfo {author} {\bibfnamefont {F.}~\bibnamefont {Meng}},\ }\href@noop {} {\bibfield  {journal} {\bibinfo  {journal} {physica status solidi (a)}\ }\textbf {\bibinfo {volume} {207}},\ \bibinfo {pages} {1481} (\bibinfo {year} {2010})}\BibitemShut {NoStop}%
\bibitem [{\citenamefont {Wei}\ \emph {et~al.}(2011{\natexlab{a}})\citenamefont {Wei}, \citenamefont {Deng}, \citenamefont {Chu}, \citenamefont {Mao}, \citenamefont {Lei},\ and\ \citenamefont {Hu}}]{Ref26}%
  \BibitemOpen
  \bibfield  {author} {\bibinfo {author} {\bibfnamefont {X.-P.}\ \bibnamefont {Wei}}, \bibinfo {author} {\bibfnamefont {J.-B.}\ \bibnamefont {Deng}}, \bibinfo {author} {\bibfnamefont {S.-B.}\ \bibnamefont {Chu}}, \bibinfo {author} {\bibfnamefont {G.-Y.}\ \bibnamefont {Mao}}, \bibinfo {author} {\bibfnamefont {T.}~\bibnamefont {Lei}},\ and\ \bibinfo {author} {\bibfnamefont {X.-R.}\ \bibnamefont {Hu}},\ }\href@noop {} {\bibfield  {journal} {\bibinfo  {journal} {Journal of Magnetism and Magnetic Materials}\ }\textbf {\bibinfo {volume} {323}},\ \bibinfo {pages} {185} (\bibinfo {year} {2011}{\natexlab{a}})}\BibitemShut {NoStop}%
\bibitem [{\citenamefont {Wei}\ \emph {et~al.}(2011{\natexlab{b}})\citenamefont {Wei}, \citenamefont {Deng}, \citenamefont {Chu}, \citenamefont {Mao}, \citenamefont {Hu}, \citenamefont {Yang},\ and\ \citenamefont {Hu}}]{Ref27}%
  \BibitemOpen
  \bibfield  {author} {\bibinfo {author} {\bibfnamefont {X.-P.}\ \bibnamefont {Wei}}, \bibinfo {author} {\bibfnamefont {J.-B.}\ \bibnamefont {Deng}}, \bibinfo {author} {\bibfnamefont {S.-B.}\ \bibnamefont {Chu}}, \bibinfo {author} {\bibfnamefont {G.-Y.}\ \bibnamefont {Mao}}, \bibinfo {author} {\bibfnamefont {L.-B.}\ \bibnamefont {Hu}}, \bibinfo {author} {\bibfnamefont {M.-K.}\ \bibnamefont {Yang}},\ and\ \bibinfo {author} {\bibfnamefont {X.-R.}\ \bibnamefont {Hu}},\ }\href@noop {} {\bibfield  {journal} {\bibinfo  {journal} {Computational Materials Science}\ }\textbf {\bibinfo {volume} {50}},\ \bibinfo {pages} {1175} (\bibinfo {year} {2011}{\natexlab{b}})}\BibitemShut {NoStop}%
\bibitem [{\citenamefont {Wei}\ \emph {et~al.}(2010)\citenamefont {Wei}, \citenamefont {Hu}, \citenamefont {Mao}, \citenamefont {Chu}, \citenamefont {Lei}, \citenamefont {Hu},\ and\ \citenamefont {Deng}}]{Ref28}%
  \BibitemOpen
  \bibfield  {author} {\bibinfo {author} {\bibfnamefont {X.-P.}\ \bibnamefont {Wei}}, \bibinfo {author} {\bibfnamefont {X.-R.}\ \bibnamefont {Hu}}, \bibinfo {author} {\bibfnamefont {G.-Y.}\ \bibnamefont {Mao}}, \bibinfo {author} {\bibfnamefont {S.-B.}\ \bibnamefont {Chu}}, \bibinfo {author} {\bibfnamefont {T.}~\bibnamefont {Lei}}, \bibinfo {author} {\bibfnamefont {L.-B.}\ \bibnamefont {Hu}},\ and\ \bibinfo {author} {\bibfnamefont {J.-B.}\ \bibnamefont {Deng}},\ }\href@noop {} {\bibfield  {journal} {\bibinfo  {journal} {Journal of Magnetism and Magnetic Materials}\ }\textbf {\bibinfo {volume} {322}},\ \bibinfo {pages} {3204} (\bibinfo {year} {2010})}\BibitemShut {NoStop}%
\bibitem [{\citenamefont {Mydosh}(1993)}]{Ref29}%
  \BibitemOpen
  \bibfield  {author} {\bibinfo {author} {\bibfnamefont {J.~A.}\ \bibnamefont {Mydosh}},\ }\href@noop {} {}\ (\bibinfo  {publisher} {CRC Press},\ \bibinfo {year} {1993})\BibitemShut {NoStop}%
\bibitem [{\citenamefont {Binder}\ and\ \citenamefont {Young}(1986)}]{Ref30}%
  \BibitemOpen
  \bibfield  {author} {\bibinfo {author} {\bibfnamefont {K.}~\bibnamefont {Binder}}\ and\ \bibinfo {author} {\bibfnamefont {A.~P.}\ \bibnamefont {Young}},\ }\href@noop {} {\bibfield  {journal} {\bibinfo  {journal} {Reviews of Modern Physics}\ }\textbf {\bibinfo {volume} {58}},\ \bibinfo {pages} {801} (\bibinfo {year} {1986})}\BibitemShut {NoStop}%
\bibitem [{\citenamefont {N{\'e}el}(1948)}]{Ref31}%
  \BibitemOpen
  \bibfield  {author} {\bibinfo {author} {\bibfnamefont {L.}~\bibnamefont {N{\'e}el}},\ }\bibfield  {title} {\bibinfo {title} {Annales de physique, paris},\ }in\ \href@noop {} {\emph {\bibinfo {booktitle} {Annales de Physique Paris}}},\ Vol.~\bibinfo {volume} {3}\ (\bibinfo {year} {1948})\ pp.\ \bibinfo {pages} {137--198}\BibitemShut {NoStop}%
\bibitem [{\citenamefont {Srinivasan}\ and\ \citenamefont {Seehra}(1983)}]{Ref32}%
  \BibitemOpen
  \bibfield  {author} {\bibinfo {author} {\bibfnamefont {G.}~\bibnamefont {Srinivasan}}\ and\ \bibinfo {author} {\bibfnamefont {M.~S.}\ \bibnamefont {Seehra}},\ }\href@noop {} {\bibfield  {journal} {\bibinfo  {journal} {Physical Review B}\ }\textbf {\bibinfo {volume} {28}},\ \bibinfo {pages} {1} (\bibinfo {year} {1983})}\BibitemShut {NoStop}%
\bibitem [{\citenamefont {Thota}\ \emph {et~al.}(2022)\citenamefont {Thota}, \citenamefont {Seehra}, \citenamefont {Chowdhury}, \citenamefont {Singh}, \citenamefont {Ghosh}, \citenamefont {Jena}, \citenamefont {Pramanik}, \citenamefont {Sarkar}, \citenamefont {Rawat}, \citenamefont {Medwal} \emph {et~al.}}]{Ref33}%
  \BibitemOpen
  \bibfield  {author} {\bibinfo {author} {\bibfnamefont {S.}~\bibnamefont {Thota}}, \bibinfo {author} {\bibfnamefont {M.~S.}\ \bibnamefont {Seehra}}, \bibinfo {author} {\bibfnamefont {M.~R.}\ \bibnamefont {Chowdhury}}, \bibinfo {author} {\bibfnamefont {H.}~\bibnamefont {Singh}}, \bibinfo {author} {\bibfnamefont {S.}~\bibnamefont {Ghosh}}, \bibinfo {author} {\bibfnamefont {S.~K.}\ \bibnamefont {Jena}}, \bibinfo {author} {\bibfnamefont {P.}~\bibnamefont {Pramanik}}, \bibinfo {author} {\bibfnamefont {T.}~\bibnamefont {Sarkar}}, \bibinfo {author} {\bibfnamefont {R.~S.}\ \bibnamefont {Rawat}}, \bibinfo {author} {\bibfnamefont {R.}~\bibnamefont {Medwal}}, \emph {et~al.},\ }\href@noop {} {\bibfield  {journal} {\bibinfo  {journal} {Physical Review B}\ }\textbf {\bibinfo {volume} {106}},\ \bibinfo {pages} {134418} (\bibinfo {year} {2022})}\BibitemShut {NoStop}%
\bibitem [{\citenamefont {Ehrenberg}\ \emph {et~al.}(1995)\citenamefont {Ehrenberg}, \citenamefont {Wltschek}, \citenamefont {Weitzel}, \citenamefont {Trouw}, \citenamefont {Buettner}, \citenamefont {Kroener},\ and\ \citenamefont {Fuess}}]{Ref34}%
  \BibitemOpen
  \bibfield  {author} {\bibinfo {author} {\bibfnamefont {H.}~\bibnamefont {Ehrenberg}}, \bibinfo {author} {\bibfnamefont {G.}~\bibnamefont {Wltschek}}, \bibinfo {author} {\bibfnamefont {H.}~\bibnamefont {Weitzel}}, \bibinfo {author} {\bibfnamefont {F.}~\bibnamefont {Trouw}}, \bibinfo {author} {\bibfnamefont {J.}~\bibnamefont {Buettner}}, \bibinfo {author} {\bibfnamefont {T.}~\bibnamefont {Kroener}},\ and\ \bibinfo {author} {\bibfnamefont {H.}~\bibnamefont {Fuess}},\ }\href@noop {} {\bibfield  {journal} {\bibinfo  {journal} {Physical Review B}\ }\textbf {\bibinfo {volume} {52}},\ \bibinfo {pages} {9595} (\bibinfo {year} {1995})}\BibitemShut {NoStop}%
\bibitem [{\citenamefont {Kumar}\ \emph {et~al.}(2020)\citenamefont {Kumar}, \citenamefont {Selvan}, \citenamefont {Vasylechko}, \citenamefont {Saravanan},\ and\ \citenamefont {Seehra}}]{Ref35}%
  \BibitemOpen
  \bibfield  {author} {\bibinfo {author} {\bibfnamefont {N.~R.}\ \bibnamefont {Kumar}}, \bibinfo {author} {\bibfnamefont {R.~K.}\ \bibnamefont {Selvan}}, \bibinfo {author} {\bibfnamefont {L.}~\bibnamefont {Vasylechko}}, \bibinfo {author} {\bibfnamefont {P.}~\bibnamefont {Saravanan}},\ and\ \bibinfo {author} {\bibfnamefont {M.~S.}\ \bibnamefont {Seehra}},\ }\href@noop {} {\bibfield  {journal} {\bibinfo  {journal} {Physica B: Condensed Matter}\ }\textbf {\bibinfo {volume} {599}},\ \bibinfo {pages} {412460} (\bibinfo {year} {2020})}\BibitemShut {NoStop}%
\bibitem [{\citenamefont {de~Almeida}\ and\ \citenamefont {Thouless}(1978)}]{Ref36}%
  \BibitemOpen
  \bibfield  {author} {\bibinfo {author} {\bibfnamefont {J.~R.}\ \bibnamefont {de~Almeida}}\ and\ \bibinfo {author} {\bibfnamefont {D.~J.}\ \bibnamefont {Thouless}},\ }\href@noop {} {\bibfield  {journal} {\bibinfo  {journal} {Journal of Physics A: Mathematical and General}\ }\textbf {\bibinfo {volume} {11}},\ \bibinfo {pages} {983} (\bibinfo {year} {1978})}\BibitemShut {NoStop}%
\bibitem [{\citenamefont {Kroder}\ \emph {et~al.}(2019)\citenamefont {Kroder}, \citenamefont {Manna}, \citenamefont {Kriegner}, \citenamefont {Sukhanov}, \citenamefont {Liu}, \citenamefont {Borrmann}, \citenamefont {Hoser}, \citenamefont {Gooth}, \citenamefont {Schnelle}, \citenamefont {Inosov} \emph {et~al.}}]{Ref37}%
  \BibitemOpen
  \bibfield  {author} {\bibinfo {author} {\bibfnamefont {J.}~\bibnamefont {Kroder}}, \bibinfo {author} {\bibfnamefont {K.}~\bibnamefont {Manna}}, \bibinfo {author} {\bibfnamefont {D.}~\bibnamefont {Kriegner}}, \bibinfo {author} {\bibfnamefont {A.}~\bibnamefont {Sukhanov}}, \bibinfo {author} {\bibfnamefont {E.}~\bibnamefont {Liu}}, \bibinfo {author} {\bibfnamefont {H.}~\bibnamefont {Borrmann}}, \bibinfo {author} {\bibfnamefont {A.}~\bibnamefont {Hoser}}, \bibinfo {author} {\bibfnamefont {J.}~\bibnamefont {Gooth}}, \bibinfo {author} {\bibfnamefont {W.}~\bibnamefont {Schnelle}}, \bibinfo {author} {\bibfnamefont {D.~S.}\ \bibnamefont {Inosov}}, \emph {et~al.},\ }\href@noop {} {\bibfield  {journal} {\bibinfo  {journal} {Physical Review B}\ }\textbf {\bibinfo {volume} {99}},\ \bibinfo {pages} {174410} (\bibinfo {year} {2019})}\BibitemShut {NoStop}%
\bibitem [{\citenamefont {Pakhira}\ \emph {et~al.}(2016)\citenamefont {Pakhira}, \citenamefont {Mazumdar}, \citenamefont {Ranganathan}, \citenamefont {Giri},\ and\ \citenamefont {Avdeev}}]{Ref38}%
  \BibitemOpen
  \bibfield  {author} {\bibinfo {author} {\bibfnamefont {S.}~\bibnamefont {Pakhira}}, \bibinfo {author} {\bibfnamefont {C.}~\bibnamefont {Mazumdar}}, \bibinfo {author} {\bibfnamefont {R.}~\bibnamefont {Ranganathan}}, \bibinfo {author} {\bibfnamefont {S.}~\bibnamefont {Giri}},\ and\ \bibinfo {author} {\bibfnamefont {M.}~\bibnamefont {Avdeev}},\ }\href@noop {} {\bibfield  {journal} {\bibinfo  {journal} {Physical Review B}\ }\textbf {\bibinfo {volume} {94}},\ \bibinfo {pages} {104414} (\bibinfo {year} {2016})}\BibitemShut {NoStop}%
\bibitem [{\citenamefont {Khorwal}\ \emph {et~al.}(2022)\citenamefont {Khorwal}, \citenamefont {Dash}, \citenamefont {Kumar}, \citenamefont {Lukoyanov}, \citenamefont {Shreder}, \citenamefont {Bitla}, \citenamefont {Vasundhara}, \citenamefont {Patra} \emph {et~al.}}]{Ref39}%
  \BibitemOpen
  \bibfield  {author} {\bibinfo {author} {\bibfnamefont {A.~K.}\ \bibnamefont {Khorwal}}, \bibinfo {author} {\bibfnamefont {S.}~\bibnamefont {Dash}}, \bibinfo {author} {\bibfnamefont {A.}~\bibnamefont {Kumar}}, \bibinfo {author} {\bibfnamefont {A.}~\bibnamefont {Lukoyanov}}, \bibinfo {author} {\bibfnamefont {E.}~\bibnamefont {Shreder}}, \bibinfo {author} {\bibfnamefont {Y.}~\bibnamefont {Bitla}}, \bibinfo {author} {\bibfnamefont {M.}~\bibnamefont {Vasundhara}}, \bibinfo {author} {\bibfnamefont {A.~K.}\ \bibnamefont {Patra}}, \emph {et~al.},\ }\href@noop {} {\bibfield  {journal} {\bibinfo  {journal} {Journal of Magnetism and Magnetic Materials}\ }\textbf {\bibinfo {volume} {546}},\ \bibinfo {pages} {168752} (\bibinfo {year} {2022})}\BibitemShut {NoStop}%
\bibitem [{\citenamefont {Alvarez}\ \emph {et~al.}(1991)\citenamefont {Alvarez}, \citenamefont {Alegra},\ and\ \citenamefont {Colmenero}}]{Ref40}%
  \BibitemOpen
  \bibfield  {author} {\bibinfo {author} {\bibfnamefont {F.}~\bibnamefont {Alvarez}}, \bibinfo {author} {\bibfnamefont {A.}~\bibnamefont {Alegra}},\ and\ \bibinfo {author} {\bibfnamefont {J.}~\bibnamefont {Colmenero}},\ }\href@noop {} {\bibfield  {journal} {\bibinfo  {journal} {Physical Review B}\ }\textbf {\bibinfo {volume} {44}},\ \bibinfo {pages} {7306} (\bibinfo {year} {1991})}\BibitemShut {NoStop}%
\bibitem [{\citenamefont {Bhattacharyya}\ \emph {et~al.}(2011)\citenamefont {Bhattacharyya}, \citenamefont {Giri},\ and\ \citenamefont {Majumdar}}]{Ref41}%
  \BibitemOpen
  \bibfield  {author} {\bibinfo {author} {\bibfnamefont {A.}~\bibnamefont {Bhattacharyya}}, \bibinfo {author} {\bibfnamefont {S.}~\bibnamefont {Giri}},\ and\ \bibinfo {author} {\bibfnamefont {S.}~\bibnamefont {Majumdar}},\ }\href@noop {} {\bibfield  {journal} {\bibinfo  {journal} {Physical Review B}\ }\textbf {\bibinfo {volume} {83}},\ \bibinfo {pages} {134427} (\bibinfo {year} {2011})}\BibitemShut {NoStop}%
\bibitem [{\citenamefont {Bag}\ \emph {et~al.}(2018)\citenamefont {Bag}, \citenamefont {Baral},\ and\ \citenamefont {Nath}}]{Ref42}%
  \BibitemOpen
  \bibfield  {author} {\bibinfo {author} {\bibfnamefont {P.}~\bibnamefont {Bag}}, \bibinfo {author} {\bibfnamefont {P.}~\bibnamefont {Baral}},\ and\ \bibinfo {author} {\bibfnamefont {R.}~\bibnamefont {Nath}},\ }\href@noop {} {\bibfield  {journal} {\bibinfo  {journal} {Physical Review B}\ }\textbf {\bibinfo {volume} {98}},\ \bibinfo {pages} {144436} (\bibinfo {year} {2018})}\BibitemShut {NoStop}%
\bibitem [{\citenamefont {Khorwal}\ \emph {et~al.}(2024{\natexlab{a}})\citenamefont {Khorwal}, \citenamefont {Vishvakarma}, \citenamefont {Dash}, \citenamefont {Bitla}, \citenamefont {Vasundhara},\ and\ \citenamefont {Patra}}]{Ref43}%
  \BibitemOpen
  \bibfield  {author} {\bibinfo {author} {\bibfnamefont {A.~K.}\ \bibnamefont {Khorwal}}, \bibinfo {author} {\bibfnamefont {S.}~\bibnamefont {Vishvakarma}}, \bibinfo {author} {\bibfnamefont {S.}~\bibnamefont {Dash}}, \bibinfo {author} {\bibfnamefont {Y.}~\bibnamefont {Bitla}}, \bibinfo {author} {\bibfnamefont {M.}~\bibnamefont {Vasundhara}},\ and\ \bibinfo {author} {\bibfnamefont {A.~K.}\ \bibnamefont {Patra}},\ }in\ \href@noop {} {\emph {\bibinfo {booktitle} {AIP Conference Proceedings}}},\ Vol.\ \bibinfo {volume} {2995}\ (\bibinfo {organization} {AIP Publishing},\ \bibinfo {year} {2024})\BibitemShut {NoStop}%
\bibitem [{\citenamefont {Chu}\ \emph {et~al.}(1994)\citenamefont {Chu}, \citenamefont {Kenning},\ and\ \citenamefont {Orbach}}]{Ref44}%
  \BibitemOpen
  \bibfield  {author} {\bibinfo {author} {\bibfnamefont {D.}~\bibnamefont {Chu}}, \bibinfo {author} {\bibfnamefont {G.}~\bibnamefont {Kenning}},\ and\ \bibinfo {author} {\bibfnamefont {R.}~\bibnamefont {Orbach}},\ }\href@noop {} {\bibfield  {journal} {\bibinfo  {journal} {Physical Review Letters}\ }\textbf {\bibinfo {volume} {72}},\ \bibinfo {pages} {3270} (\bibinfo {year} {1994})}\BibitemShut {NoStop}%
\bibitem [{\citenamefont {Khorwal}\ \emph {et~al.}(2024{\natexlab{b}})\citenamefont {Khorwal}, \citenamefont {Saha}, \citenamefont {Lukoyanov},\ and\ \citenamefont {Patra}}]{Ref43a}%
  \BibitemOpen
  \bibfield  {author} {\bibinfo {author} {\bibfnamefont {A.~K.}\ \bibnamefont {Khorwal}}, \bibinfo {author} {\bibfnamefont {S.}~\bibnamefont {Saha}}, \bibinfo {author} {\bibfnamefont {A.~V.}\ \bibnamefont {Lukoyanov}},\ and\ \bibinfo {author} {\bibfnamefont {A.~K.}\ \bibnamefont {Patra}},\ }\href@noop {} {\bibfield  {journal} {\bibinfo  {journal} {The Journal of Chemical Physics}\ }\textbf {\bibinfo {volume} {160}} (\bibinfo {year} {2024}{\natexlab{b}})}\BibitemShut {NoStop}%
\bibitem [{\citenamefont {Sun}\ \emph {et~al.}(2003)\citenamefont {Sun}, \citenamefont {Salamon}, \citenamefont {Garnier},\ and\ \citenamefont {Averback}}]{Ref45}%
  \BibitemOpen
  \bibfield  {author} {\bibinfo {author} {\bibfnamefont {Y.}~\bibnamefont {Sun}}, \bibinfo {author} {\bibfnamefont {M.}~\bibnamefont {Salamon}}, \bibinfo {author} {\bibfnamefont {K.}~\bibnamefont {Garnier}},\ and\ \bibinfo {author} {\bibfnamefont {R.}~\bibnamefont {Averback}},\ }\href@noop {} {\bibfield  {journal} {\bibinfo  {journal} {Physical Review Letters}\ }\textbf {\bibinfo {volume} {91}},\ \bibinfo {pages} {167206} (\bibinfo {year} {2003})}\BibitemShut {NoStop}%
\bibitem [{\citenamefont {Sasaki}\ \emph {et~al.}(2005)\citenamefont {Sasaki}, \citenamefont {J{\"o}nsson}, \citenamefont {Takayama},\ and\ \citenamefont {Mamiya}}]{Ref46}%
  \BibitemOpen
  \bibfield  {author} {\bibinfo {author} {\bibfnamefont {M.}~\bibnamefont {Sasaki}}, \bibinfo {author} {\bibfnamefont {P.}~\bibnamefont {J{\"o}nsson}}, \bibinfo {author} {\bibfnamefont {H.}~\bibnamefont {Takayama}},\ and\ \bibinfo {author} {\bibfnamefont {H.}~\bibnamefont {Mamiya}},\ }\href@noop {} {\bibfield  {journal} {\bibinfo  {journal} {Physical Review B}\ }\textbf {\bibinfo {volume} {71}},\ \bibinfo {pages} {104405} (\bibinfo {year} {2005})}\BibitemShut {NoStop}%
\bibitem [{\citenamefont {Fisher}\ and\ \citenamefont {Huse}(1988)}]{Ref47}%
  \BibitemOpen
  \bibfield  {author} {\bibinfo {author} {\bibfnamefont {D.~S.}\ \bibnamefont {Fisher}}\ and\ \bibinfo {author} {\bibfnamefont {D.~A.}\ \bibnamefont {Huse}},\ }\href@noop {} {\bibfield  {journal} {\bibinfo  {journal} {Physical Review B}\ }\textbf {\bibinfo {volume} {38}},\ \bibinfo {pages} {373} (\bibinfo {year} {1988})}\BibitemShut {NoStop}%
\bibitem [{\citenamefont {McMillan}(1984)}]{Ref48}%
  \BibitemOpen
  \bibfield  {author} {\bibinfo {author} {\bibfnamefont {W.}~\bibnamefont {McMillan}},\ }\href@noop {} {\bibfield  {journal} {\bibinfo  {journal} {Journal of Physics C: Solid State Physics}\ }\textbf {\bibinfo {volume} {17}},\ \bibinfo {pages} {3179} (\bibinfo {year} {1984})}\BibitemShut {NoStop}%
\bibitem [{\citenamefont {Lefloch}\ \emph {et~al.}(1992)\citenamefont {Lefloch}, \citenamefont {Hammann}, \citenamefont {Ocio},\ and\ \citenamefont {Vincent}}]{Ref49}%
  \BibitemOpen
  \bibfield  {author} {\bibinfo {author} {\bibfnamefont {F.}~\bibnamefont {Lefloch}}, \bibinfo {author} {\bibfnamefont {J.}~\bibnamefont {Hammann}}, \bibinfo {author} {\bibfnamefont {M.}~\bibnamefont {Ocio}},\ and\ \bibinfo {author} {\bibfnamefont {E.}~\bibnamefont {Vincent}},\ }\href@noop {} {\bibfield  {journal} {\bibinfo  {journal} {Europhysics Letters}\ }\textbf {\bibinfo {volume} {18}},\ \bibinfo {pages} {647} (\bibinfo {year} {1992})}\BibitemShut {NoStop}%
\bibitem [{\citenamefont {Mukherjee}\ \emph {et~al.}(2012)\citenamefont {Mukherjee}, \citenamefont {Garg},\ and\ \citenamefont {Gupta}}]{Ref50}%
  \BibitemOpen
  \bibfield  {author} {\bibinfo {author} {\bibfnamefont {S.}~\bibnamefont {Mukherjee}}, \bibinfo {author} {\bibfnamefont {A.}~\bibnamefont {Garg}},\ and\ \bibinfo {author} {\bibfnamefont {R.}~\bibnamefont {Gupta}},\ }\href@noop {} {\bibfield  {journal} {\bibinfo  {journal} {Applied Physics Letters}\ }\textbf {\bibinfo {volume} {100}} (\bibinfo {year} {2012})}\BibitemShut {NoStop}%
\bibitem [{\citenamefont {Anand}\ \emph {et~al.}(2012)\citenamefont {Anand}, \citenamefont {Adroja},\ and\ \citenamefont {Hillier}}]{Ref51}%
  \BibitemOpen
  \bibfield  {author} {\bibinfo {author} {\bibfnamefont {V.}~\bibnamefont {Anand}}, \bibinfo {author} {\bibfnamefont {D.}~\bibnamefont {Adroja}},\ and\ \bibinfo {author} {\bibfnamefont {A.}~\bibnamefont {Hillier}},\ }\href@noop {} {\bibfield  {journal} {\bibinfo  {journal} {Physical Review B}\ }\textbf {\bibinfo {volume} {85}},\ \bibinfo {pages} {014418} (\bibinfo {year} {2012})}\BibitemShut {NoStop}%
\bibitem [{\citenamefont {Gopal}(2012)}]{Ref52}%
  \BibitemOpen
  \bibfield  {author} {\bibinfo {author} {\bibfnamefont {E.}~\bibnamefont {Gopal}},\ }\href@noop {} {}\ (\bibinfo  {publisher} {Springer Science \& Business Media},\ \bibinfo {year} {2012})\BibitemShut {NoStop}%
\bibitem [{\citenamefont {Li}\ \emph {et~al.}(1998)\citenamefont {Li}, \citenamefont {Shiokawa}, \citenamefont {Homma}, \citenamefont {Uesawa}, \citenamefont {D{\"o}nni}, \citenamefont {Suzuki}, \citenamefont {Haga}, \citenamefont {Yamamoto}, \citenamefont {Honma},\ and\ \citenamefont {{\=O}nuki}}]{Ref53}%
  \BibitemOpen
  \bibfield  {author} {\bibinfo {author} {\bibfnamefont {D.}~\bibnamefont {Li}}, \bibinfo {author} {\bibfnamefont {Y.}~\bibnamefont {Shiokawa}}, \bibinfo {author} {\bibfnamefont {Y.}~\bibnamefont {Homma}}, \bibinfo {author} {\bibfnamefont {A.}~\bibnamefont {Uesawa}}, \bibinfo {author} {\bibfnamefont {A.}~\bibnamefont {D{\"o}nni}}, \bibinfo {author} {\bibfnamefont {T.}~\bibnamefont {Suzuki}}, \bibinfo {author} {\bibfnamefont {Y.}~\bibnamefont {Haga}}, \bibinfo {author} {\bibfnamefont {E.}~\bibnamefont {Yamamoto}}, \bibinfo {author} {\bibfnamefont {T.}~\bibnamefont {Honma}},\ and\ \bibinfo {author} {\bibfnamefont {Y.}~\bibnamefont {{\=O}nuki}},\ }\href@noop {} {\bibfield  {journal} {\bibinfo  {journal} {Physical Review B}\ }\textbf {\bibinfo {volume} {57}},\ \bibinfo {pages} {7434} (\bibinfo {year} {1998})}\BibitemShut {NoStop}%
\bibitem [{\citenamefont {Sangeetha}\ \emph {et~al.}(2019)\citenamefont {Sangeetha}, \citenamefont {Wang}, \citenamefont {Smirnov}, \citenamefont {Smetana}, \citenamefont {Mudring}, \citenamefont {Johnson}, \citenamefont {Tanatar}, \citenamefont {Prozorov},\ and\ \citenamefont {Johnston}}]{Ref54}%
  \BibitemOpen
  \bibfield  {author} {\bibinfo {author} {\bibfnamefont {N.~S.}\ \bibnamefont {Sangeetha}}, \bibinfo {author} {\bibfnamefont {L.-L.}\ \bibnamefont {Wang}}, \bibinfo {author} {\bibfnamefont {A.~V.}\ \bibnamefont {Smirnov}}, \bibinfo {author} {\bibfnamefont {V.}~\bibnamefont {Smetana}}, \bibinfo {author} {\bibfnamefont {A.-V.}\ \bibnamefont {Mudring}}, \bibinfo {author} {\bibfnamefont {D.}~\bibnamefont {Johnson}}, \bibinfo {author} {\bibfnamefont {M.}~\bibnamefont {Tanatar}}, \bibinfo {author} {\bibfnamefont {R.}~\bibnamefont {Prozorov}},\ and\ \bibinfo {author} {\bibfnamefont {D.}~\bibnamefont {Johnston}},\ }\href@noop {} {\bibfield  {journal} {\bibinfo  {journal} {Physical Review B}\ }\textbf {\bibinfo {volume} {100}},\ \bibinfo {pages} {094447} (\bibinfo {year} {2019})}\BibitemShut {NoStop}%
\bibitem [{\citenamefont {Tien}\ \emph {et~al.}(2000)\citenamefont {Tien}, \citenamefont {Feng}, \citenamefont {Wur},\ and\ \citenamefont {Lu}}]{Ref55}%
  \BibitemOpen
  \bibfield  {author} {\bibinfo {author} {\bibfnamefont {C.}~\bibnamefont {Tien}}, \bibinfo {author} {\bibfnamefont {C.~H.}\ \bibnamefont {Feng}}, \bibinfo {author} {\bibfnamefont {C.~S.}\ \bibnamefont {Wur}},\ and\ \bibinfo {author} {\bibfnamefont {J.~J.}\ \bibnamefont {Lu}},\ }\href@noop {} {\bibfield  {journal} {\bibinfo  {journal} {Physical Review B}\ }\textbf {\bibinfo {volume} {61}},\ \bibinfo {pages} {12151} (\bibinfo {year} {2000})}\BibitemShut {NoStop}%
\bibitem [{\citenamefont {Dash}\ \emph {et~al.}(2020)\citenamefont {Dash}, \citenamefont {Lukoyanov}, \citenamefont {Mishra}, \citenamefont {Rasi}, \citenamefont {Gangineni}, \citenamefont {Vasundhara}, \citenamefont {Patra} \emph {et~al.}}]{Ref56}%
  \BibitemOpen
  \bibfield  {author} {\bibinfo {author} {\bibfnamefont {S.}~\bibnamefont {Dash}}, \bibinfo {author} {\bibfnamefont {A.}~\bibnamefont {Lukoyanov}}, \bibinfo {author} {\bibfnamefont {D.}~\bibnamefont {Mishra}}, \bibinfo {author} {\bibfnamefont {U.~M.}\ \bibnamefont {Rasi}}, \bibinfo {author} {\bibfnamefont {R.}~\bibnamefont {Gangineni}}, \bibinfo {author} {\bibfnamefont {M.}~\bibnamefont {Vasundhara}}, \bibinfo {author} {\bibfnamefont {A.~K.}\ \bibnamefont {Patra}}, \emph {et~al.},\ }\href@noop {} {\bibfield  {journal} {\bibinfo  {journal} {Journal of Magnetism and Magnetic Materials}\ }\textbf {\bibinfo {volume} {513}},\ \bibinfo {pages} {167205} (\bibinfo {year} {2020})}\BibitemShut {NoStop}%
\end{thebibliography}%

\end{document}